\documentclass{aa}  

\usepackage{graphicx}
\usepackage{xcolor}
\usepackage{txfonts}
\usepackage{placeins}
\usepackage{hyperref}
\newcommand{\micron}{$\rm\mu$m}

\begin{document}

   \title{The disk of the eruptive protostar V900~Mon\thanks{Based on observations collected at the European Southern Observatory under programs 092.C-0513(A), 094.C-0476(B), 0104.C-0607(A) and 0106.C-0510(A).}}
   \subtitle{a MATISSE/VLTI and MUSE/VLT perspective}

 \author{F.~Lykou \inst{1,2}
          \and
          P.~\'Abrah\'am\inst{1,2,3,4}
          \and
        F.~Cruz-S\'aenz de Miera \inst{1,2,5}
        \and
          J.~Varga\inst{1,6}
        \and
        \'A.~K\'osp\'al\inst{1,2,3,7}
          \and
          J.~Bouwman\inst{7}
        \and
        L.~Chen \inst{1,2}
        \and
        S.~Kraus \inst{8}
        \and
        M.~L.~Sitko\inst{9,10,11}
        \and
        R.~W.~Russell\inst{12,13,11}
        \and
        M.~Pikhartova\inst{14}
          }

   \institute{Konkoly Observatory, HUN-REN Research Centre for Astronomy and Earth Sciences (CSFK), Konkoly-Thege Mikl\'os \'ut 15-17, 1121 Budapest, Hungary 
        \and
        CSFK, MTA Centre of Excellence, Budapest, Konkoly Thege Mikl\'os \'ut 15-17., H-1121, Hungary
         \and
         ELTE E\"otv\"os Lor\'and University, Institute of Physics, P\'azm\'any P\'eter s\'et\'any 1/A, 1117 Budapest, Hungary
         \and
         Department of Astrophysics, University of Vienna, T\"urkenschanzstrasse 17, 1180 Vienna, Austria
         \and
         Institut de Recherche en Astrophysique et Plan\'etologie, Universit\'e de Toulouse, UT3-PS, OMP, CNRS, 9 av. du Colonel-Roche, 31028, Toulouse Cedex 4, France
         \and
         Leiden Observatory, Leiden University, P.O. Box 9513, NL2300, RA Leiden, The Netherlands
         \and
         Max Planck Institute for Astronomy, K\"onigstuhl 17, D-69117 Heidelberg, Germany
         \and
         Astrophysics Group, Department of Physics and Astronomy, University of Exeter, Stocker Road, Exeter EX4 4QL, UK
         \and
         Space Science Institute, 4765 Walnut St., Suite B, Boulder, CO 80301, USA
         \and
        Department of Physics, University of Cincinnati, Cincinnati, OH 45221, USA
        \and
                Visiting Astronomer, NASA Infrared Telescope Facility
        \and
        Observations Unlimited, 1417 11th Street, Manhattan Beach, CA 90266-6107, USA
        \and
        The Aerospace Corporation, P.O. Box 92957, Los Angeles, CA 90009-2957, USA
        \and
        Prague, 15000, Czech Republic 
		}

   \date{Received Dec 20, 2022; accepted Nov XX, 2023}

 
  \abstract
   {} 
  {
  In this work, we study the silicate dust content in the disk of one of the youngest eruptive stars, V900 Mon, at the highest angular resolution probing down to the inner 10~au of said disk, and study the historical evolution of the system traced in part by a newly discovered emission clump.
  }
   {We performed high-angular resolution mid-infrared interferometric observations of V900~Mon with MATISSE/VLTI with a spatial coverage ranging from 38-m to 130-m baselines, and compared them to archival MIDI/VLTI data. We also mined and re-analyzed archival optical and infrared photometry of the star to study its long-term evolution since its eruption in the 1990s. We complemented our findings with integral field spectroscopy data from MUSE/VLT.
   }
   {The MATISSE/VLTI data suggest a radial variation of the silicate feature in the dusty disk, whereby at large spatial scales ($\geq10$~au) the protostellar disk's emission is dominated by large-sized ($\geq1$\micron) silicate grains, while at smaller spatial scales and closer to the star ($\leq5$~au), silicate emission is absent suggesting self-shielding. We propose that the self-shielding may be the result of small dust grains at the base of the collimated CO outflow previously detected by ALMA. A newly discovered knot in the MUSE/VLT data, located at a projected distance approximately 27,000 au from the star, is co-aligned with the molecular gas outflow at a P.A. of 250\degr ($\pm5$\degr) consistent with the position angle and inclination of the disk. The knot is seen in emission in H$\alpha$, [\ion{N}{ii}], and the [\ion{S}{II}] doublet and its kinematic age is about 5150 years. This ejected material could originate from a previous eruption.
   }
  {}


   \maketitle
%

\section{Introduction}

Mass accretion is an important factor in the process of star formation. In the case of low-mass young stellar objects (YSOs), and during their transition from Class I to Class II, circumstellar material is accreted onto a protostar through its circumstellar disk. Between evolutionary stages, the mass accretion rate can vary by a factor of 100--1000 or higher, especially if the star goes through eruptive phases  \citep[e.g.,][]{hartmann1998}. It is a matter of debate whether eruptive phases are common in all low-mass YSOs, or these occur only in a unique sub-class of objects \citep[e.g., ][]{fischer2022}. The most well-known category of such events are the FU Orionis-type (FUor) eruptions \citep{herbig1966, herbig1977, hartmann1996,audard2014}.

FUor eruptions are perhaps the most powerful events during star formation. The outbursts are due to episodically highly increased mass accretion (as high as $\sim$10$^{-4}$ M$_{\odot}$\,yr$^{-1}$), leading to extraordinarily high bolometric luminosity (up to 500 L$_{\odot}$), which may last for a few decades. The archetype of the class, FU~Orionis, was already known to have a protoplanetary disk \citep{1985ApJ...299..462H}. More recently, studies using high-angular-resolution imaging techniques have found it to be rather compact compared to typical protoplanetary disks \citep[e.g.,][]{perez2020,liu2021,lykou2022} . One of the most recent members of the FUor class is V900~Mon whose eruption occurred sometime between 1950 and 2010 \citep{reipurth2012}. It appears that the eruption is still on-going based on photometric monitoring in the optical and infrared \citep{samus2011, semkov2021}. The star is surrounded by an intricate reflection nebula discovered by an amateur astronomer \citep{thommes2011}.


V900 Mon is located at a distance of $1227^{+130}_{-111}$~pc \citep[{\it Gaia} DR3; ][]{2021AJ....161..147B} in the vicinity of the Galactic Plane ($b\approx -2.5^{o}$). The Infrared Astronomical Satellite (IRAS) dust extinction maps suggest an interstellar extinction $A_V \approx 6.4$ mag in this direction \citep{sf2011} but this measurement may be affected by the strong far-infrared emission of nearby region LBN1022. \citet{car2022} used the 5780\AA\ and 6614\AA\ diffuse interstellar bands (DIBs) to derive the local, interstellar extinction to V900~Mon at $A_V \approx 2.8\pm 0.4$ mag. Young stellar objects are usually deeply embedded in the dust clouds from which they form, therefore it is expected that the extinction toward V900 Mon might be higher when accounting for circumstellar matter. \citet{connelley2018} suggest an extinction of $A_V=13.5\pm 2$~mag, similar to \citet{reipurth2012}, by comparing the near-infrared spectrum of V900~Mon to that of FU~Ori. This is not surprising since the higher value of \citet{connelley2018} contains both circumstellar and interstellar extinction.

Atacama Large Millimeter/submillimeter Array (ALMA) observations of the molecular gas (carbon monoxide, CO) around the star suggest a bi-conical spatial distribution with an opening angle of 70\degr\ for the blue-shifted \element[][12]{CO} emission \citep{takami2019}. This low-velocity component extends westward up to 5\arcsec\ from the protostar, while the bulk of the emission arises within a region approximately 0.5\arcsec\ in radius. The direction of said western outflow is oriented at roughly 250\degr\ east-of-north. Bi-conical, molecular-emission outflows typically have wide opening angles and low-velocity components (e.g., $<100$~km~s$^{-1}$) as opposed to collimated jets that show narrow opening angles with high-velocity shock-excited gas emission (e.g., $\ge 100$~km~s$^{-1}$). \citet{takami2019} hypothesized that the low-velocity outflow seen by ALMA can be formed by a collimated jet interacting with the surrounding gas.

Collimated jets have been found in just a few of the already known eruptive stars. Well-known examples are V346~Nor \citep{kospal2020b} as evidenced from forbidden-line emission in the optical and the near-infrared (e.g., [\ion{O}{i}], \ion{[S}{ii]}, \ion{[Fe}{ii]}), and the binary system Z~CMa \citep{zcmajet}. \citet{reipurth2012} obtained integral field unit (IFU) spectroscopic observations of V900~Mon with the Near-Infrared Integral Field Spectrometer (NIFS) on Gemini North telescope, but they did not find signatures of jet-like emission within the instrumental 3\arcsec$\times$3\arcsec\ field-of-view (FOV).

The intricate molecular emission can be associated with the reflection nebula around V900~Mon. The nebula is very bright only on the western side of the protostar, while if there is any emission eastward, this is very faint as is the \element[][12]{CO} red-shifted emission (e.g., Fig.~2 in \citet{reipurth2012} against Fig.~2 in \citet{takami2019}). This is consistent with a disk seen pole-on, as material behind the disk can be obscured. The complex reflection nebula displays ``helicoidal''-like features as seen in broadband imaging, as for example in the $r'$-band maps of \citet{reipurth2012} or in archival data from the VST Photometric H$\alpha$ Survey of the Southern Galactic Plane and Bulge \citep[VPHAS+, ][]{vphas} in the red, and UKIRT Infrared Deep Sky Survey \citep[UKIDSS, ][]{ukidss} in the near-infrared.

According to \citet{takami2019}, a 1.3-mm continuum-emission component was marginally resolved at $200\times150$ milliarcseconds (mas) by ALMA. The authors estimate deconvolved sizes for the disk at about $67\times58$ mas. This translates to approximately $82\times71$ au at the {\it Gaia} distance. A re-analysis of those data sets by \citet{kospal2021} found that the disk position angle (PA) is $169\pm73$ degrees and the inclination is $i=28\pm20$ degrees\footnote{Based on Gaussian fit estimates. The authors also mention visibility-fit estimates at slightly higher values, PA$=181^{+27}_{-21}$ and $i=34^{+15}_{-16}$ degrees, which are in agreement.}, suggesting the disk is nearly pole-on in our line of sight. \citet{kospal2021} give a lower limit for the disk mass at 0.01~M$_\sun$ when utilizing the bolometric luminosity of \citet{reipurth2012}, however their {\sc radmc-3d} radiative transfer model predicts a higher disk mass at 0.3~M$_\sun$ making this one of the most massive FUor disks.

Eruptive phenomena cause strong feedback on the surrounding circumstellar environment. Among the different physical, chemical, and mineralogical processes, the crystallization of amorphous silicates is one that was always anticipated in the literature but has never been actually observed in FUors \citep[e.g.,][]{quanz2006}. Interestingly, crystallization via thermal annealing at T$>$1000 K has been detected in the less powerful and shorter outburst of EX Lup \citep{abraham2009}. \citet{kospal2020} analyzed VLT Imager and Spectrometer for mid Infrared (VISIR/VLT) mid-infrared spectra of V900~Mon that indicated the presence of large-sized silicate grains in the disk, but could not provide a definitive answer on the presence of crystalline silicates at spatial scales $\leq200$~mas.

An initial study of eruptive stars at high-angular resolution in the mid-infrared was performed by \citet{varga2018} with the MID-infrared Interferometric instrument \citep[MIDI; ][]{midi} of the Very Large Telescope Interferometer (VLTI). Recent technological developments with the introduction of the Multi AperTure mid-Infrared SpectroScopic Experiment \citep[MATISSE; ][]{matisse3,matisse}, which is the imaging interferometric instrument of the VLTI that operates in the mid-infrared (3 -- 13 \micron), allow the simultaneous observations of eruptive star disks at superior $uv$-coverage. Following the observations of the archetype FU~Ori \citep{lykou2022}, we have employed MATISSE for an in-depth analysis of the disk of V900~Mon.

We also follow up on the detection of a collimated molecular outflow by ALMA \citep{takami2019} with wide-field, optical IFU observations of V900~Mon with the Multi Unit Spectroscopic Explorer \citep[MUSE; ][]{muse1} on the VLT.

A description of the observing program follows below (Sect.~\ref{sec:obs}), while the observational results are presented in Section~\ref{sec:results}. Interpretations of said results can be found in Section \ref{sec:discussion}, and these are followed by our conclusions in the last section.

\section{Observations and methodology}\label{sec:obs}

\subsection{MATISSE}
We obtained snapshot observations with MATISSE/VLTI in $L$ and $N$ bands (0104.C-0607(A); P.I.: P.~\'Abrah\'am). The source was observed at 2 epochs with the 8.2-m Unit Telescope (UT) array on 2019 December 11 and 2020 January 9. The observation log can be found in Table~\ref{tab:logvlti}. 

In this work, we will use the generic term ``baseline''  as opposed to the more accurate term ``projected baseline'' that takes into account its projection -- length and/or position angle -- on the sky. The baseline coverage offered by the UTs was between 38 and 130 meters, and the respective angular resolutions ranged at $\sim$ 3--9 mas in the $L$-band and $\sim$ 8--26 mas in the $N$-band. We opted for the UT array due to the source's faintness (correlated flux, $F_{\rm corr, N} < 5$~Jy). Although MATISSE offers simultaneous observations in the $M$-band, these were rendered useless in this case, since the target's correlated $M$-band flux was lower than the designated limits\footnote{\url{https://www.eso.org/sci/facilities/paranal/instruments/matisse/inst.html}} for the UTs. Hence, the $M$-band is not used in this work. The FOV of MATISSE for the UTs is the size of the pinhole spatial filter, that is $1.5\lambda/D \approx\,$0.13\arcsec\ at 3.5 \micron\ and $2\lambda/D \approx\,$0.5\arcsec\ at 10 \micron , where $D=8.2$ m is the diameter of a single UT.

The observations were obtained in the low-spectral resolution mode for all spectral bands ($R=\lambda/\Delta\lambda \sim30$). Due to the lack of a hybrid calibrator for both bands, we opted for a CAL-SCI-CAL sequence, whereby each science exposure was bracketed by a calibrator suitable for the $N$ and the $L$ band: HD47667 (K2III; \diameter\ =2.61 mas) and HD59381 (K4/5III; \diameter\ = 2.33 mas). The calibrator diameters were obtained from \citet{jsdc}. The data sets were reduced, processed, and calibrated with the standard MATISSE DRS pipeline (v.1.7.5) and the wrapper Python tools\footnote{\url{https://gitlab.oca.eu/MATISSE/tools}}. Flux calibration was performed with a customized tool as in \citet{varga2021}.


\subsection{MUSE}

In addition to the interferometric data, we report here MUSE/VLT observations of the circumstellar environment of V900~Mon and its
reflection nebula. These were obtained on 2021 January 24 (program ID: 106.21KL; P.I.: F.~Cruz-Sáenz de Miera) using the WFM-AO-N mode. The ``wide field mode'' (WFM) provides a field of view covering $1'\times1'$ and a pixel size of $0.2'' \times 0.2''$. MUSE's nominal mode (N) has a spectral coverage from 4700\AA\ to 9300\AA\ with a resolving power between 1770 and 3590. The Na Notch filter of the adaptive optics (AO)
system creates a discontinuity in the spectrograph at 5800--5970\AA. The observations were executed in six 175\,s exposures, resulting in a total integration time of 17.5\,minutes. Each exposure after the first one was dithered by 1$''$ and rotated by 90$^{\circ}$. 

The observations were reduced using the MUSE pipeline (v.2.8.4) by ESO \citep{muse2}, resulting in a fully calibrated and combined MUSE data cube with the instrumental signature and local sky background removed. Further analysis, such as extraction of stellar spectra and integrated images, was performed with designated IFU tools: {\tt mpdaf}\footnote{\url{https://mpdaf.readthedocs.io/}} and {\sc QFitsView}\footnote{\url{https://www.mpe.mpg.de/~ott/QFitsView/}}.

\begin{figure}[btp]
    \centering
    \includegraphics[width=\columnwidth]{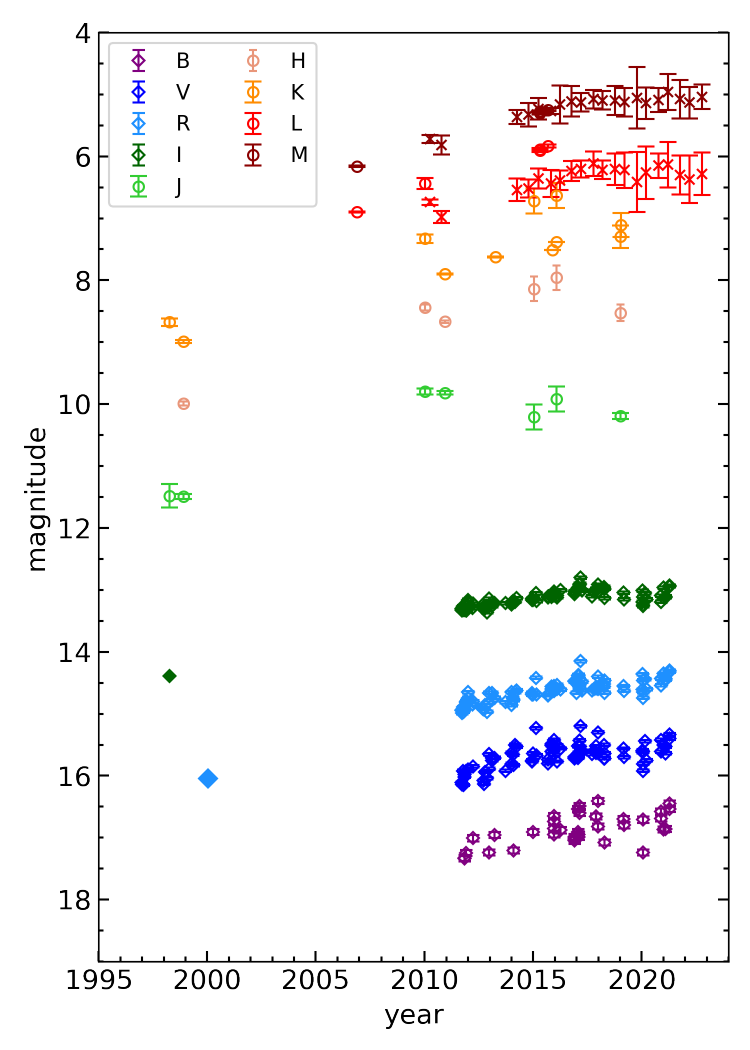}
    \caption{Evolution of V900~Mon. For a description of the data, see main text in Sections~\ref{irphot} and \ref{sec:evol}.
    }
    \label{fig:allphot}
\end{figure}

\subsection{IRTF spectroscopy}\label{irtf}
We obtained near-infrared (NIR) spectra using the SpeX spectrograph \citep{rayner03} on 2015 January 15 and 2016 January 25-26 (cf. Table~\ref{tab:logphot}). The spectra were recorded using the Echelle grating in both short-wavelength mode (SXD, 0.8–2.4 \micron)  and long wavelength mode (LXD, 2.3–5.4 \micron) using a 0.8\arcsec\ slit. The spectra were reduced, corrected for telluric absorption and flux calibrated against a number of A0V calibrator stars, using the {\sc Spextool} data reduction package (for a detailed description of the methods used, we refer to \citet{vacca03} and \citet{cushing04}).

Photometry was obtained using the SpeX guide camera in the $K$ band on 4 nights between 2016 January 22-26.  The entrance aperture had a radius of 1.9\arcsec\ (15 pixels on the camera) and the subtracted sky background was determined using an annulus of 2.4\arcsec\ to 3.6\arcsec\ (20-30 pixels).

The mid-infrared (MIR) spectra were obtained with the Aerospace Corporation’s Broad-band Array Spectrograph System (BASS) on 2015 January 14. BASS uses a cold beamsplitter to separate the light into two separate wavelength regimes. The short wavelength beam includes light from 2.9--6 \micron, while the long-wavelength beam covers 6--13.5 \micron. Each beam is dispersed onto a 58-element Blocked Impurity Band linear array, thus allowing for simultaneous coverage of the spectrum from 2.9--13.5 \micron. The spectral resolution $R$ is wavelength-dependent, ranging from about 30 to 125 over each of the two wavelength regions \citep{hackwell90}. The (fixed) circular entrance aperture was 4\arcsec. The standard star $\alpha$ Tau (K5+III), which was observed at a similar airmass to V900 Mon, was used for absolute photometric calibration.

The SpeX spectra from 2015 January 15 were scaled (due to light loss at the slit for the SpeX observations) to the flux-calibrated BASS data. The SpeX spectra from 2016 January 25-26 were scaled to the mean of the $K$-band photometry on 2016 January 22-26 (Table~\ref{tab:nirphot}).


\subsection{Infrared Photometry}\label{irphot}

We obtained near-infrared $JHK$ images using three instruments between 2012 and 2019: CAIN-III on the 1.52-m Telescopio Carlos Sanchez (TCS) at Teide Observatory in Tenerife, Spain (P.I.: P.~\'Abrah\'am; detector specifications: 256x256 pixels, 1\arcsec/pix), and A Novel Dual Imaging CAMera (ANDICAM) on the 1.3m Small and Moderate Aperture Research Telescope System (SMARTS) instrument located at Cerro Tololo Interamerican Observatory (CTIO) in Chile (P.I.: \'A.~K\'osp\'al; detector specifications: 512x512 pixels, 0.27\arcsec/pix). We also mined archival data from REMIR on the Rapid Eye Mount telescope \citep[REM; ][]{rem} in La Silla Observatory, Chile (P.I: M.~Cur\'e; detector specifications: 512x512 pixels, 1.23\arcsec/pix). The log of observations is presented in Table~\ref{tab:logphot}. 

Observations were performed in a 5-point dither pattern in order to enable proper sky subtraction. Exposure times of individual frames ranged between 0.5 and 6 sec. The images were reduced using our custom-made IDL\footnote{\url{https://www.l3harrisgeospatial.com/Software-Technology/IDL}} routines. Data reduction steps included sky subtraction, flat-fielding, registration, and co-adding exposures by dither position and filter. To calibrate our photometry, we used the Two Micron All Sky Survey (2MASS) catalog \citep{2mass}. The instrumental magnitudes of the target and typically 5-15 good-quality 2MASS stars in the field were extracted using an aperture radius of 2\arcsec\ and a sky annulus between 3.6\arcsec\ and 4.8\arcsec\ in all filters. The annulus was intentionally placed inside the extended nebulosity surrounding the star, in an attempt to remove the nebula's contribution, that finally turned out to be only a minor factor in the photometry. We determined a constant offset between the instrumental and the 2MASS magnitudes and used it for calibrating V900~Mon. The results are listed in Table~\ref{tab:nirphot} and Figure~\ref{fig:allphot}.

\subsection{MIDI}\label{midi}

V900~Mon was observed with MIDI/VLTI on 2013 December 21 and 2015 January 7 (program IDs: 092.C-0513(A), 094.C-0476(B); PI: M.~Cur\'e) with the UT2-UT3 (47 m) and UT1-UT4 (130 m) baselines, respectively. Both observations were obtained in average seeing conditions (0.8\arcsec\ -- 1.1\arcsec) however the latter suffered from short atmospheric coherence time (Table~\ref{tab:logvlti}).

The first data set was reduced and analyzed for the MIDI Atlas project\footnote{\url{https://konkoly.hu/MIDI_atlas/}} \citep{varga2018}. The second data set was obtained with the use of an external fringe tracker, namely the Phase-Referenced Imaging and Micro-arcsecond Astrometry Fringe Tracking Unit A (PRIMA FSU-A), operating in the $K$ band \citep{prima}. Only the correlated flux is recorded in this instrumental mode, therefore no photometry was obtained for this run. This data set was reduced with MIDI's EWS pipeline using calibrators HD36673 (F0Ib, \diameter\ = 1.69 mas) and HD44951 (K2III, \diameter\ = 1.73 mas). The calibrator properties were extracted from \citet{jsdc}. 

The MIDI correlated spectra shown here (Fig.~\ref{allMATdata}) are the mean of the individual spectra from each calibrator. The typical uncertainty for MIDI is about 10\% \citep{Chesneau2007}. Since both MIDI observations were obtained with baselines identical to those of MATISSE, we compare these data sets' correlated fluxes against the MATISSE data in Section~\ref{sec:mine}.

\begin{figure}[btp]
    \centering
        \includegraphics[width=\columnwidth]{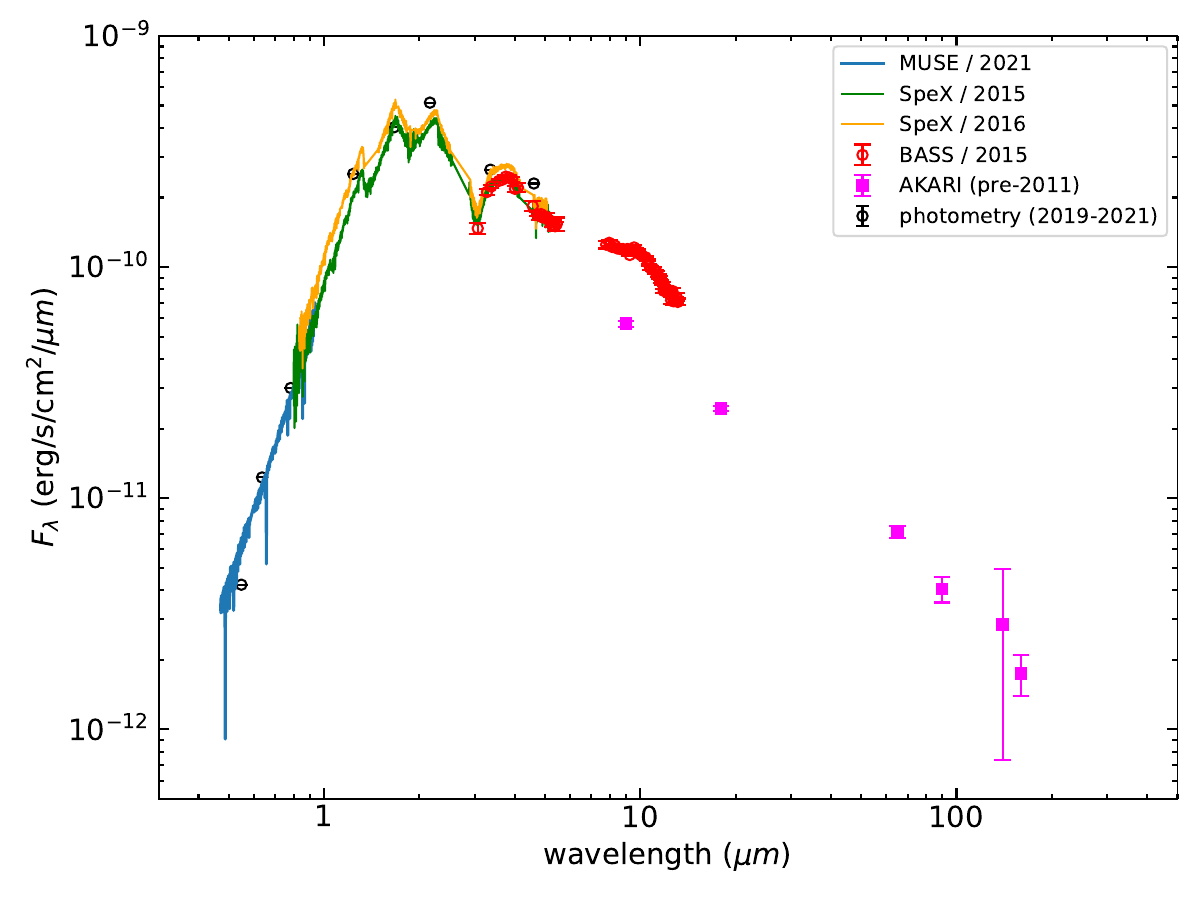}
    \caption{The spectral energy distribution of V900~Mon from the optical to the far-infrared. The two absorption lines seen in the MUSE spectrum are the H$\beta$ and H$\alpha$ lines. The SED has not been corrected for extinction.  }
    \label{fig:SED}
\end{figure}

\section{Results}\label{sec:results}

\subsection{The evolution of V900~Mon}\label{sec:evol}


The optical light curve of V900~Mon is shown in Fig.~\ref{fig:allphot}. The archival data were extracted from \citet{semkov2021} in $BVR_cI_c$ filters including earlier Deep Near Infrared Survey of the Southern Sky (DENIS) $I$-band photometry.

Figure~\ref{fig:allphot} also shows the infrared light curve of V900~Mon in $JHKL$ and $M$ bands. It comprises of our own photometry (Sect.~\ref{irphot}) and archival data from: DENIS ($JK_s$), 2MASS ($JHK_s$), {\it Spitzer} (3.6 \& 4.5 \micron), the Wide-field Infrared Survey Explorer \citep[WISE, 2010-2012; ][]{wise}, the Near-Earth Object Wide-field Infrared Survey Explorer Reactivation Mission \citep[NEOWISE, post-2014; ][]{neowise2}, \citet[][$JHK$]{samus2011}, \citet[][$JHKL'M'$]{reipurth2012}, \citet[][$L'M'$]{atel}, and \citet[][$K$]{connelley2018}. The WISE and NEOWISE data are binned per 180 days and have been corrected for saturation following the instrument's documentation\footnote{\url{https://wise2.ipac.caltech.edu/docs/release/neowise/expsup/sec2_1civa.html}}. Differences between the WISE/NEOWISE and ground-based photometry (e.g., 2011 and 2016) can be attributed to different photometric apertures used between satellite and terrestrial observations, an over-correction for source saturation in the WISE/NEOWISE data, and differences in filters used.

As is evident in Fig.~\ref{fig:allphot} the source continues to slowly brighten in the last 20 years with a temporary fading in the period 2018-2019. The infrared lightcurve in $J$, $H$, and $K$ could suggest an overshoot in 2010 \citep[photometry by ][]{reipurth2012}, however there was no contemporaneous photometry in the optical to confirm this.

Figure.~\ref{fig:SED} shows the spectral energy distribution (SED) of V900~Mon, which is that of a typical FUor. 
 SpeX/IRTF spectra also indicate a small brightening within a single year similar to what is seen in the lightcurve (Fig.~\ref{fig:allphot}). Earlier photometry by the AKARI satellite might also indicate that the source was fainter in the mid-infrared before 2011. The SED of Fig.~\ref{fig:SED} has not been corrected for extinction.

The evolution of the source in the near-infrared and the optical is shown in the color-magnitude diagrams in Fig.~\ref{fig:cmdiag}, respectively. The $V-R$ vs. $V$ color-magnitude diagram of the \citet{semkov2021} optical photometry suggests that  a decrease in the extinction may have contributed to the brightening of V900 Mon after 2012. However, the $R-I$ vs. $R$ diagram (Fig.\ref{fig:cmdiag}, bottom) indicates that the source became bluer after the 2018/9 dimming.


\begin{figure}[htbp]
\centering
		\includegraphics[width=0.99\columnwidth]{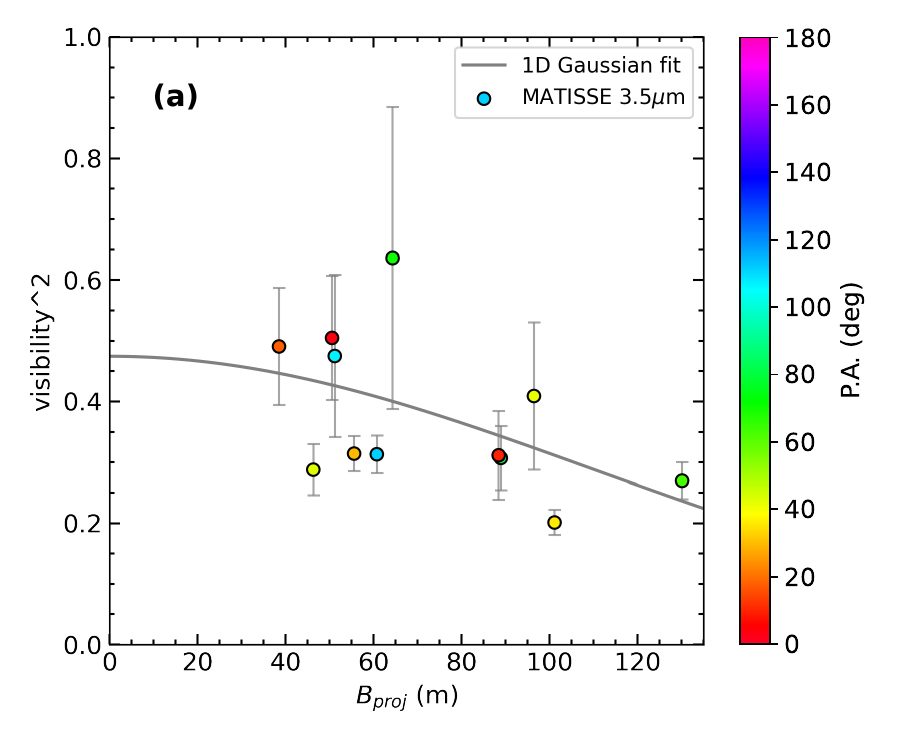}\\
		\includegraphics[width=0.99\columnwidth]{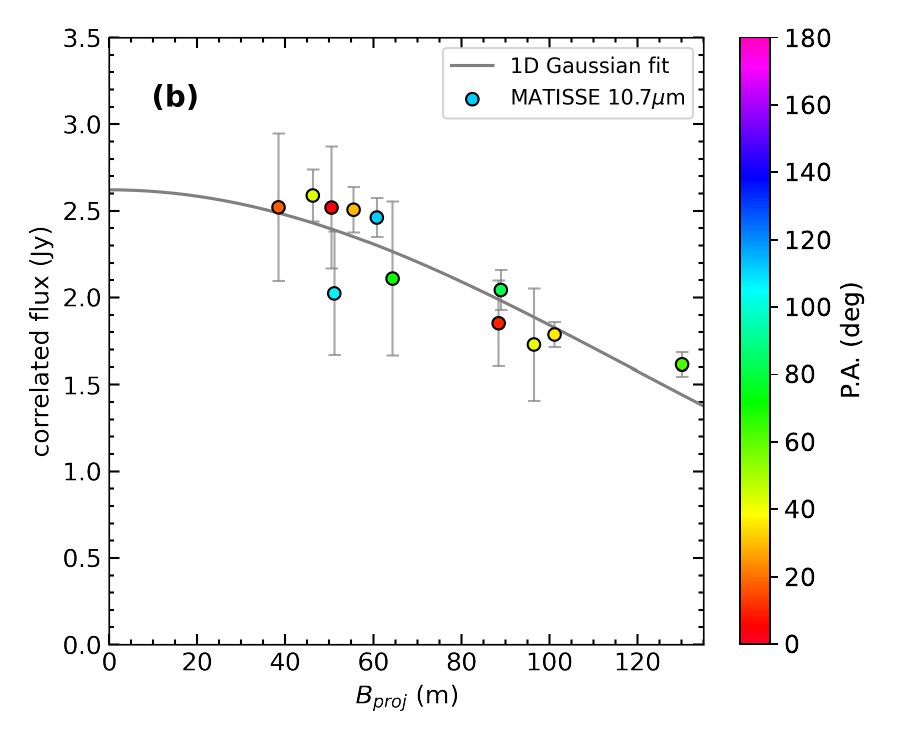}
\caption{Centro-symmetric model fits (1D Gaussian distributions) to the monochromatic squared visibilities at 3.5\micron\ (top panel), and to the monochromatic correlated fluxes at 10.7\micron\ (bottom panel).
	}
\label{allMATdata}
\end{figure}

\begin{figure}[hbtp]
    \centering
\includegraphics[width=0.9\columnwidth]{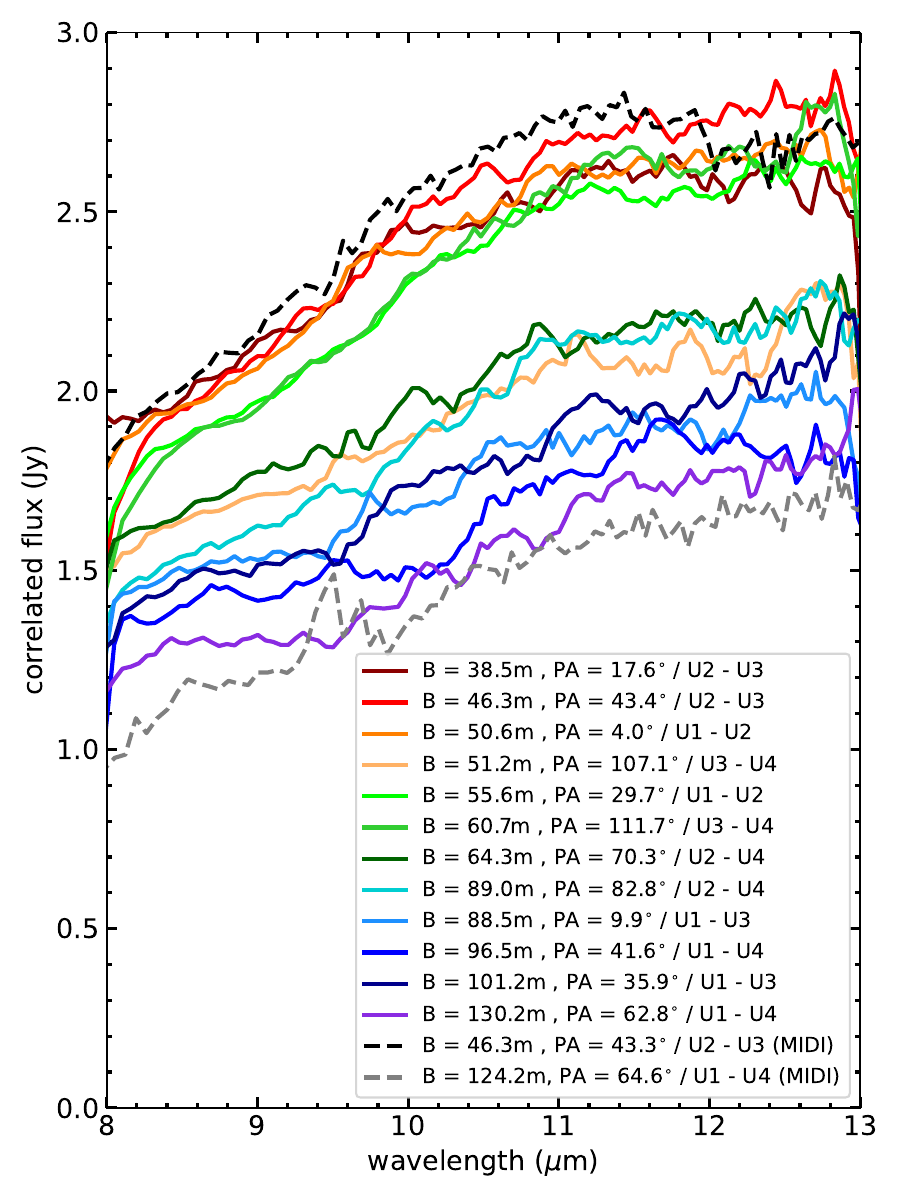}\\
\includegraphics[width=0.9\columnwidth]{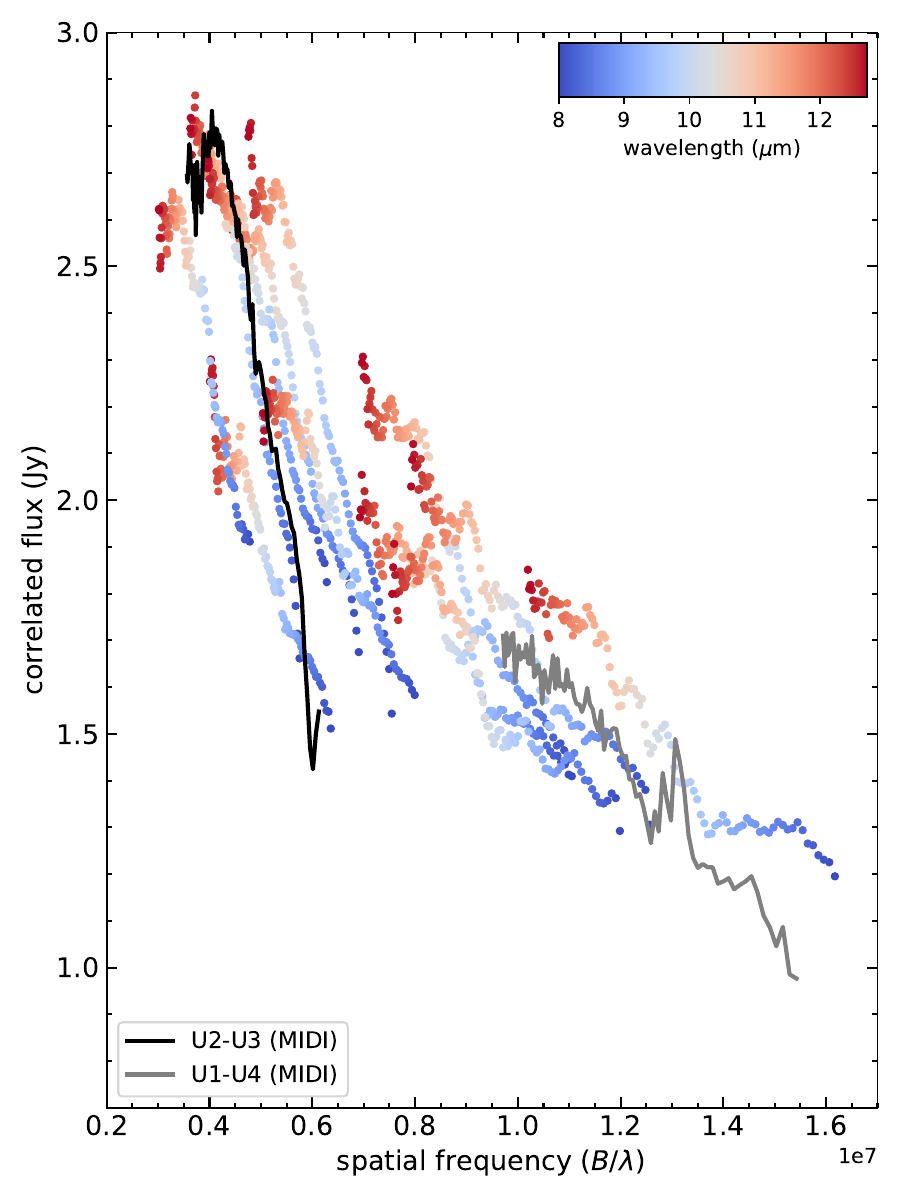}
    \caption{MATISSE and MIDI correlated fluxes. {\it Top panel:} each MATISSE correlated spectrum is color-coded by individual baseline, and in the
legend, we specify the length and position angle of each one. The apparent decrease in flux levels with increasing baseline length from top to bottom is indicative of a radial variation of the emitting region. Also shown here are the MIDI spectra (black and gray dashed lines) that follow a similar pattern. {\it Bottom panel:} the MATISSE correlated fluxes are plotted vs. spatial frequency and they are color-coded with respect to wavelength (as indicated in the colorbar). The MIDI data are shown for reference (black and gray lines).
    }
    \label{fig:allfcorr}
\end{figure}

\subsection{MATISSE}\label{sec:matresults}

Typically in young stellar objects -- including FUors -- the inner accretion disk extends to a distance of $\leq$1 au from the star, while the outer passive disk extends further outward up to $\sim 100$~au  \citep[e.g., ][and references therein]{dulle2010,hartmann2016}. The hot inner disk emits at wavelengths below 3 \micron, while the cooler outer disk emits mainly at longer wavelengths (e.g., $N$-band and sub-millimeter).

The MATISSE $L$- and $N$-band data suggest a partially-resolved source with squared visibilities\footnote{The $N$-band visibilities were not calibrated due to the low quality of the obtained photometry. In this paper we only show the $N$-band correlated fluxes.} $V^2 \geq 0.2$ at 3.5 \micron . Initial estimates of source's geometry were made by fitting 1D Gaussian functions to the MATISSE data at 3.5 \micron\ and 10.7 \micron\ (Fig.~\ref{allMATdata}a,b). The full width at half maximum (FWHM) for each band is $2.4\pm0.2$ and $7.0\pm0.2$ mas, respectively. These translate to physical diameters of about 3~au and 9~au, respectively. Therefore, the inner accretion disk is confined within an 1.5~au radius, while the majority of the emission of the dusty disk in the $N$-band originates within a radius of roughly 10~au.

The MATISSE closure phases, shown indicatively in Figs.~\ref{fig:Lcps} and \ref{fig:Ncps}, were of poorer S/N due primarily to atmospheric variations during the observing runs. Although some closure phases are non-zero, which could suggest asymmetry, they also appear to be relatively constant but the uncertainties are quite large. As such, the closure phases are not used in this work.

The most prominent result from these MATISSE observations is the variation of the $N$-band correlated fluxes with respect to the baselines. In Figure~\ref{fig:allfcorr} (top panel), the correlated fluxes are plotted from top to bottom per increasing baseline length for each epoch indicating the difference in flux levels. The distinct decrease in correlated flux levels per baseline suggests a radial variation of the emitting region. This is corroborated by the MIDI correlated fluxes (black and gray lines in Fig.~\ref{fig:allfcorr}) that were obtained at similar baselines (cf Sect.~\ref{midi}), which bracket the MATISSE data and follow a similar pattern per baseline. This is variation is also illustrated when the MATISSE correlated spectra are shown against spatial frequency and color-coded per wavelength (bottom panel, Figure~\ref{fig:allfcorr}). We explore this variation aspect further in subsequent sections (Sect.~\ref{sec:mine}).

V900~Mon's disk was detected first in the millimeter regime by ALMA \citep{takami2019} and it was found to be oriented nearly pole-on (cf. Table~\ref{tab:alma_sizes}). Here, we attempt to estimate the disk's 2D size in the $N$ band following the same method as in \citet{varga2021}. The fitting process of a 2D Gaussian distribution resulted in a FWHM of $6.2^{+1.5}_{-0.3}$~mas at 9.5 \micron\ with a major axis PA of $158^{+3}_{-119}$ degrees, and an inclination $i=14^{+4}_{-3}$ degrees ($\chi^2 =2.4$). 
For such a small inclination, the fit of the PA was not well constrained as is evident by the large errors, because the disk itself is nearly pole-on and it is therefore difficult to constrain its ellipsoidal geometry. Literature values for the disk's inclination vary (Table~\ref{tab:alma_sizes}). Recently, \citet{kospal2021} calculated an inclination of $28\pm20$ degrees from ALMA data. Unless higher angular-resolution ALMA observations are obtained in the future, we can not conclude about a potential misalignment between the MATISSE and ALMA detections.

\begin{table*}[htbp]
    \centering
    \caption{Disk sizes (FWHM) at 1.3~mm continuum from the literature compared with MATISSE $N$-band results.}
    \begin{tabular}{l|ccccc}
    \hline
    Reference   & Major axis (mas) &   Minor axis (mas) &   P.A. (deg)  &   Inclination (deg) \\ \hline\hline
    \citet{takami2019}  &   $67\pm8$    &   $58\pm8$    &   --  &   0/60$\dagger$ \\
    \citet{hales2020}   &   $72\pm11$   &   $60\pm20$   &   $164\pm63$  &   50$\dagger$  \\
    \citet{kospal2021}*  &  $43\pm4$ & $38\pm8$ &  $169\pm73$  &   $28\pm20$   \\
    this work, MATISSE  &  $6.24^{+1.54}_{-0.29}$ & $6.05\pm1.61$ & $158^{+3}_{-119}$ & $14^{+4}_{-3}$ \\
    \hline
    \end{tabular}
    \tablefoot{($\dagger$) : model; (*) \citet{kospal2021} re-analyzed the data of \citet{takami2019}. }
    \label{tab:alma_sizes}
\end{table*}


\subsection{MUSE}\label{sec:muse}

The MUSE data are a forest of information. In this work, we explore the relationship between star, disk, and environment using gas and/or dust measurements from ALMA, MATISSE, and MUSE. A full analysis of the MUSE data sets will be presented in a forthcoming paper (F.~Cruz-S\'aenz~de~Miera et al., in preparation).

\begin{figure}
    \includegraphics[width=\columnwidth]{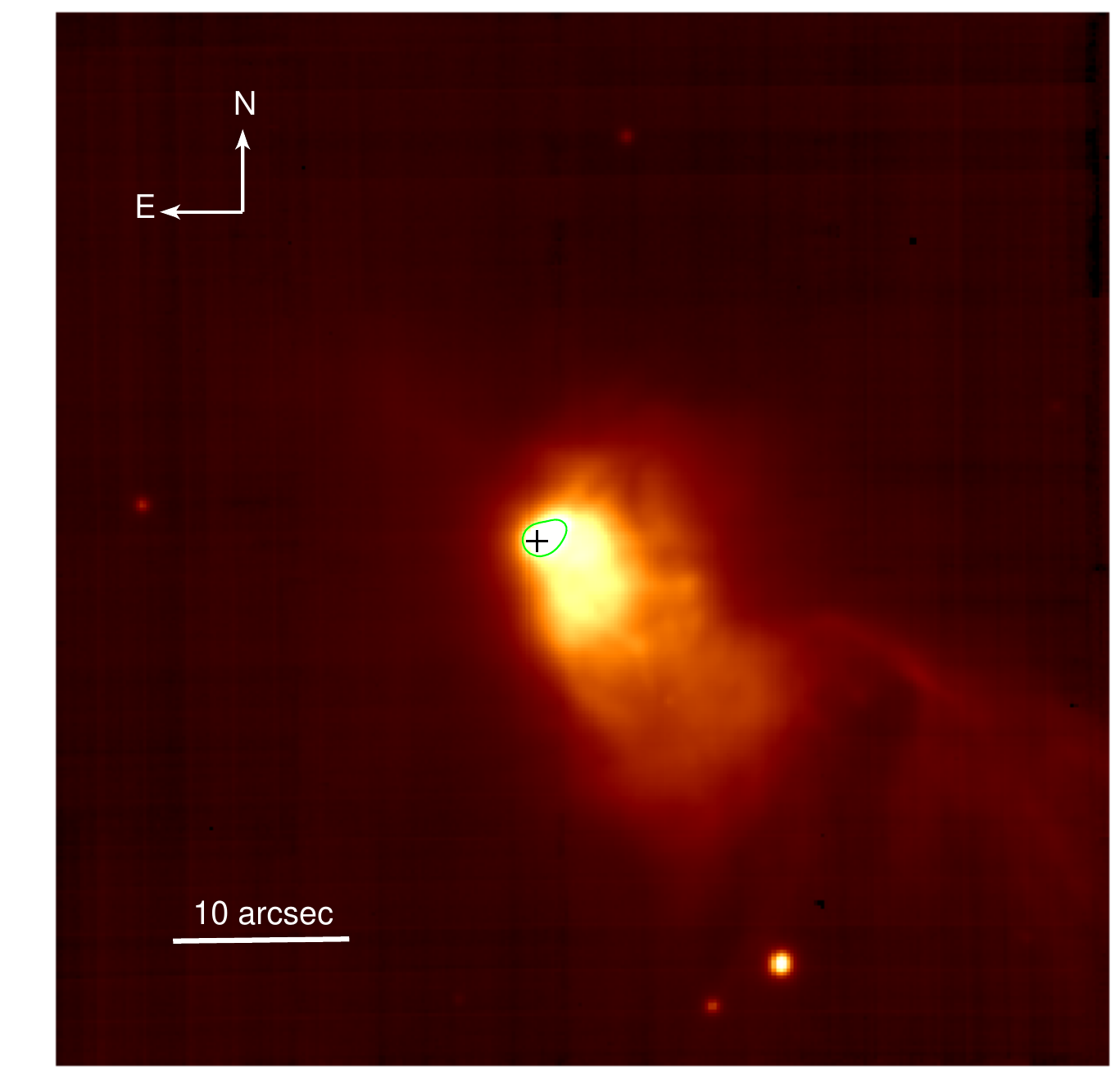}
    \caption{White image of V900~Mon and Thommes' nebula from MUSE (4700 -- 9300\AA). The cross sign marks the location of the star and the drawn green line marks the bulk of the emitting region that includes the ellipsoidal component (Sect. \ref{sec:synphot}). The field of view is 1\arcmin$\times$1\arcmin. The image has been stretched at arbitrary levels of the square root of intensity to enhance nebula features. }
    \label{fig:muse_cont_im}
\end{figure}


\subsubsection{The spectrum of the inner, circumstellar region}

Figure~\ref{fig:muse_cont_im} shows the white image of the collapsed MUSE spectral cube along the entire spectral axis (4700 -- 9300\AA). As is evident, emission from the reflection nebula (Thommes' nebula) dominates. A description of the complex nebular morphology can be found in Sect.~\ref{nebmorph} .

The star itself is not resolved, but the majority of the emission originates within a few MUSE spaxels around the {\it Gaia} position of V900~Mon. Our aim was to construct the spectrum of the inner part of the system (cf. Sect.~\ref{sec:matresults}) with minimal contribution from the nebula. Therefore, even if the physical radius of the protoplanetary disk was 100~au, the disk's angular size would be $\approx 160$~mas at the adopted {\it Gaia} distance, and as such the entire system is confined within a single MUSE spaxel. We consider this as the inner, circumstellar region or else the ``star+disk'' component.

\begin{figure*}[htbp]
    \centering
\includegraphics[width=\textwidth]{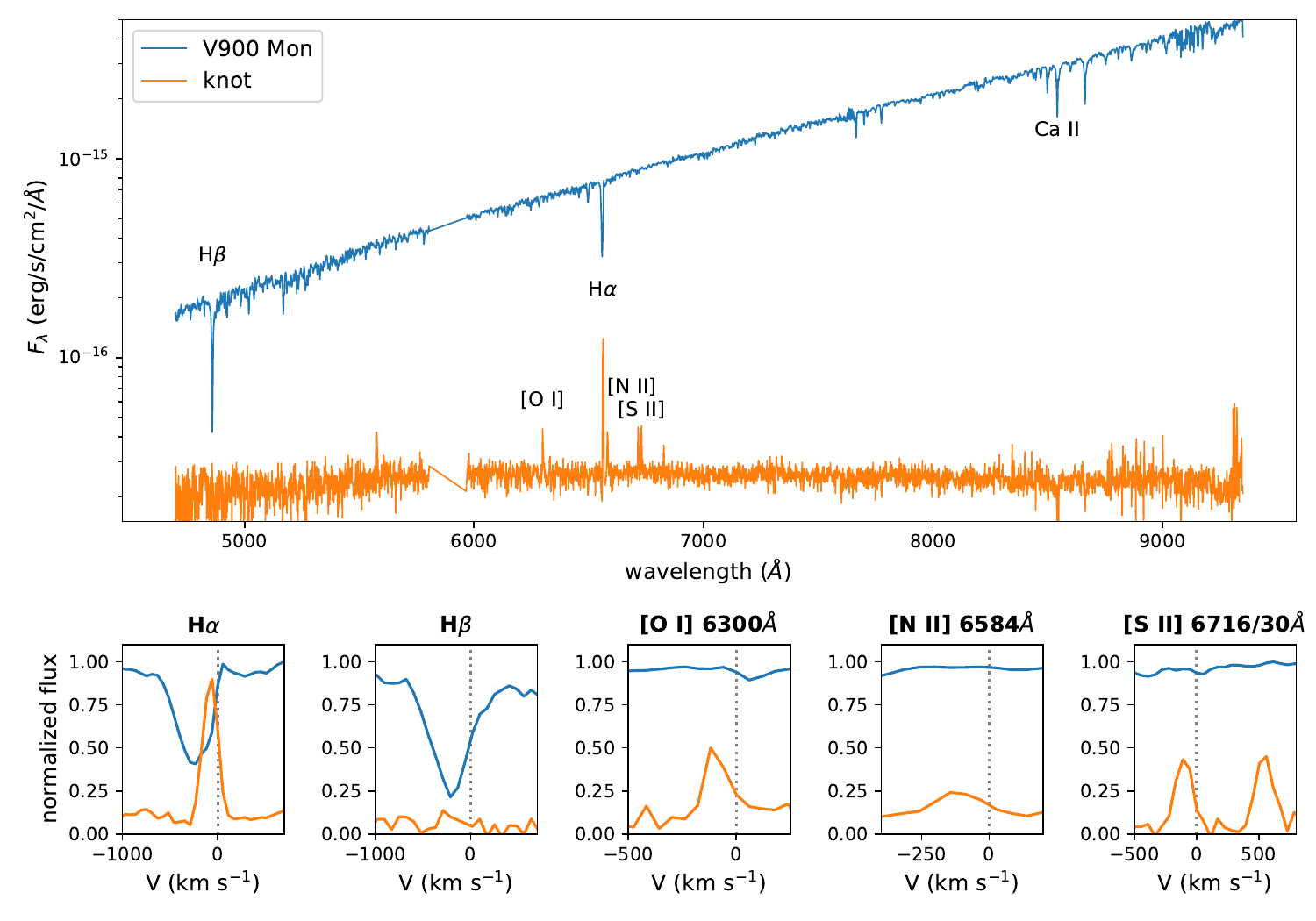}
    \caption{The MUSE spectra of V900 Mon (star+disk; blue) and the emission-line knot (orange). Prominent lines in the V900 Mon spectrum, like hydrogen Balmer lines and the \ion{Ca}{ii} triplet, are seen in absorption. In the knot spectrum, the H$\alpha$, [\ion{N}{ii}], and [\ion{S}{ii}] lines are clearly seen in emission, while the [\ion{O}{i}] line suffers from terrestrial contamination (e.g., airglow \ion{O}{i}). Both spectra are affected by skyline contamination at the blue and red ends. The spectra have not been corrected for extinction. Normalized line profiles vs. radial velocities (corrected for systemic velocity) are shown in the bottom panels.}
    \label{fig:muse_stellar_spec}
\end{figure*}

We extracted a spectrum within an 0.6\arcsec\ radius aperture at the location of V900~Mon. The spectrum (blue, Fig.~\ref{fig:muse_stellar_spec}) is typical FUor-like with lines such as H$\beta$, H$\alpha$, and the \ion{Ca}{ii} triplet, seen in absorption. All lines are blue-shifted with respect to rest wavelengths (bottom panels, Fig.~\ref{fig:muse_stellar_spec}). Typical signature lines of jet-like origin are absent (e.g., \ion{[O}{i]} at 5577, 6300, and 6364\AA\ and Fe lines). We can not compare the \ion{Na}{D} doublet to earlier results by \citet{reipurth2012}, since this falls within the spectral discontinuity of MUSE.

The Balmer lines are quite broad (width larger than 600~km~s$^{-1}$). We estimate the blue-shifted radial velocities of the Balmer lines -- after applying barycentric and heliocentric corrections -- at the deepest troughs at about $ -230\pm 10$~km\,s$^{-1}$. These velocities are slightly higher than those found by \citet{reipurth2012} for H$\alpha$, however their observations were obtained at higher spectral resolution which allowed the separation of individual peaks within the trough.


\subsubsection{Synthetic photometry}\label{sec:synphot}


When comparing the above-mentioned MUSE IFU spectrum to the near-infrared SpeX spectrum, we note a significant difference in absolute flux levels; the MUSE spectrum is fainter by a factor of thirty. We attribute this difference to the continuum contribution from the reflection nebula (Fig.~\ref{fig:muse_cont_im}), which has also been shown to emit strongly in the near-infrared \citep{reipurth2012}, since the absolute photometric calibration of the SpeX spectra was based on wide-aperture photometry. If we increase the aperture radius for the extracted MUSE spectrum to 1\arcsec, and thereby allow additional continuum emission within the aperture, then we can finally anchor the two spectra. We also note that when such a large aperture is used, the blue end of the MUSE spectrum increases, suggesting an increase in scattered light emission from the nebula.

As such, extracting synthetic photometry of the star+disk system from the MUSE data can be problematic. One-dimensional cuts vertically and horizontally through the star's position at various cube channels, indicate three sources of emission: (1) the star+disk system (within a 1\arcsec\ radius), (2) an ellipsoidal-like component adjacent to the star toward the north-west (cf.~Fig.~\ref{fig:muse_cont_im}), and (3) a large-scale source to the west and south-west (Thommes' nebula). To minimize contribution from the last two sources, we have opted to subtract the nebular emission over each channel map using a median-subtraction algorithm and an $11\times11$ pixels box. Although this technique leaves some residual emission over the nebula, it can separate the stronger emission-components, that is the star+disk and the ellipsoidal component.

Broadband-equivalent images were created from this median-subtracted spectral cube in the Johnson $V$ and Cousins $R_c$ and $I_c$ filters with {\tt mpdaf} routines. To further remove the residual contribution from the nebula to the spectrum of the circumstellar region, we clip the western section of each broadband image (i.e., west of the star), and replace it with a mirror image of the eastern section. The latter had been already ``cleaned'' of any nebular emission by the median-subtraction process. Aperture photometry with an aperture radius of 2\arcsec\ was extracted from these images with various methods to confirm consistency. The synthetic photometry in Johnson $V$, Cousins $R_c$ and Cousins $I_c$ bands is 17.34, 15.67, and 14.01 mag, respectively. The uncertainty (0.06 mag) reflects the differences between the aperture photometry algorithms used in this exercise. The MUSE photometry is used to constrain the model in Sect.~\ref{model}.


\begin{figure}[htbp]
    \centering
        \includegraphics[width=\columnwidth]{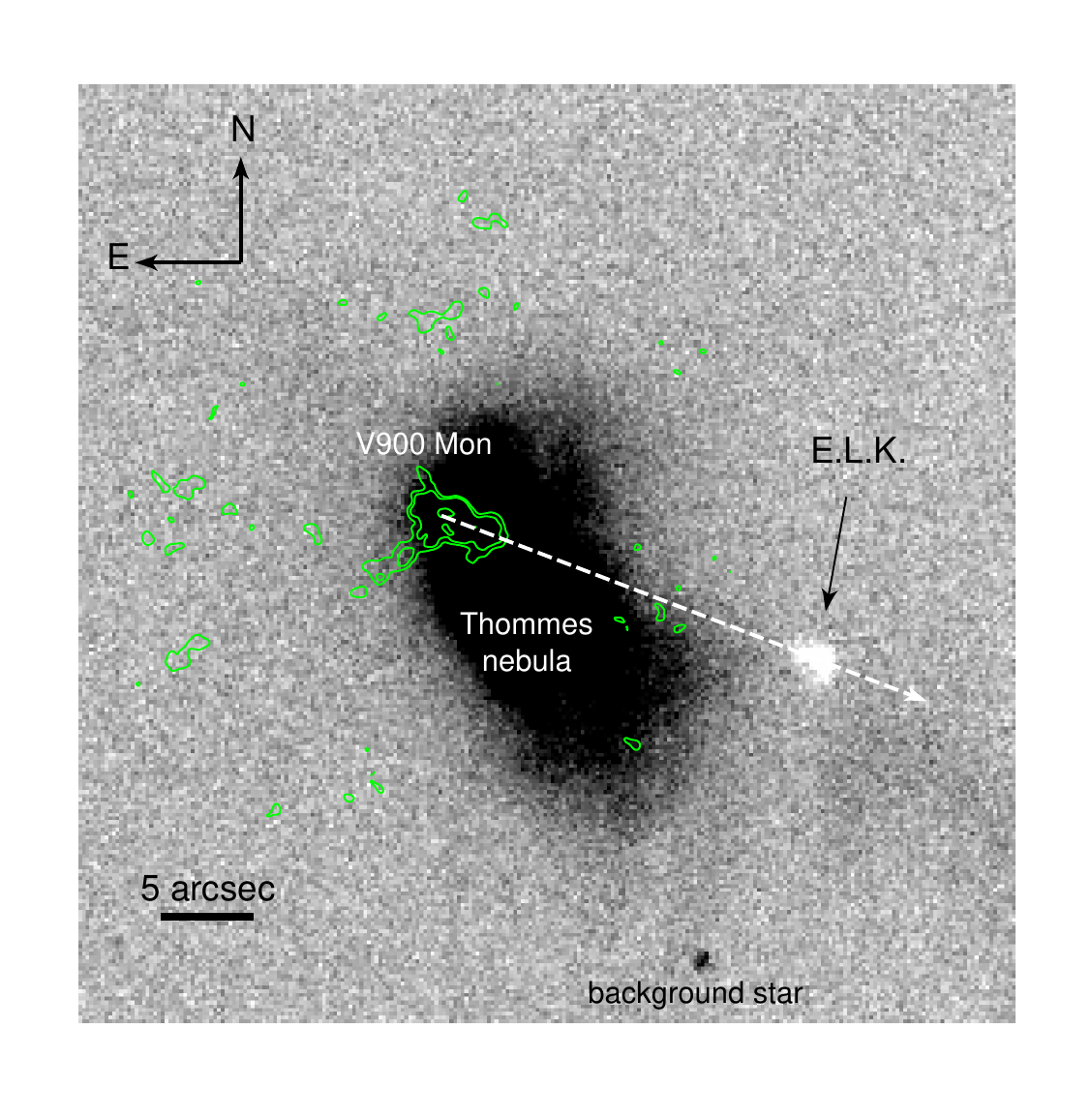}
    \caption{Continuum-subtracted $H\alpha$ linemap of V900 Mon and Thommes' nebula. The map has been linearly scaled to enhance the emission features (in white) and designate scattered light continuum from the nebula and background sources (black-shaded regions). The green contours (arbitrary levels) mark the CO (2-1) moment zero map similar to \citet{takami2019}, showing the blue-shifted emission component to the west and part of the red-shifted wide-angle lobe to the east. The emission-line knot (E.L.K.) is clearly visible and co-aligned to the blue-shifted CO outflow, as indicated by the guiding arrow (PA$\sim250$\degr\ east-of-north).  The field of view is 50\arcsec$\times$50\arcsec. }
    \label{fig:linemap}
\end{figure}

\subsubsection{Discovery of an emission-line knot}\label{knot}

A visual inspection of the continuum-subtracted line maps at the Balmer lines, as well as the typical lines suggesting jet-like signatures (e.g., [\ion{O}{i}], [\ion{S}{ii}]), revealed an emission-line knot at a distance of 22\arcsec $\pm$1\arcsec\ from the stellar position (``E.L.K.'' in Fig.~\ref{fig:linemap}). The projected distance of the knot, by adopting the {\it Gaia} distance, translates to approximately 27,000 au or 0.13 pc. 
The size of the knot is approximately 1.30\arcsec$\times$0.98\arcsec\ (2D Gaussian fit). The knot is well aligned to the collimated CO outflow detected within 5\arcsec\ from the star by \citet{takami2019}, that is within a PA of $250\pm5$ degrees east-of-north. This is perpendicular to the disk's major axis. If the disk's inclination is $i=14$\degr, then the knot's distance can be de-projected (i.e., when divided by $\sin(i)$) onto the polar axis of the disk, which translates to approximately 111,600 au or 0.54 pc.

Figure~\ref{fig:muse_stellar_spec} also shows the knot's spectrum (orange). The H$\alpha$ line is dominant while H$\beta$ is not detected (that spectral region is of lower S/N). It also shows the [\ion{O}{i}] 6300\AA\ in emission but this might be blended with terrestrial \ion{O}{i} emission, while the 6364\AA\ line is absent or it is too weak to be detected. On the other hand, the [\ion{N}{ii}] blue component at 6548\AA\ is very weak as opposed to the red component at 6584\AA . Most importantly, the [\ion{S}{ii}] doublet (6716/30\AA) is seen in emission. All of the above hints that the gas is shock-excited.

All emission lines are blue-shifted with respect to rest wavelengths with an average radial velocity (after applying barycentric and heliocentric corrections; cf. bottom panels in Fig.~\ref{fig:muse_stellar_spec}) of about $-100$~km\,s$^{-1}$. The de-projected velocity of the knot in the plane of sky is 25~km\,s$^{-1}$ and the knot's kinematic age is approximately 5150 years, assuming constant velocity. This age is far greater than the alleged post-1960s eruption of V900~Mon, however the true motion of the knot can only be revealed with future IFU observations to determine whether the gas is in expansion.

We used {\tt mpdaf} routines to calculate integrated fluxes for each emission line, and the {\tt PyNeb} software \citep{pyneb} for line diagnostics. From the [\ion{S}{ii}] 6716/30\AA\ line flux ratio 
we estimated an electron density $n_e = 950\pm100$\,cm$^{-3}$ assuming $T_e=10,000$~K. From the Balmer line ratio and assuming Case B recombination as per \citet{osterbrock2006}, we deduced an extinction of $A_V=3.4\pm0.2$ mag for typical Milky Way values ($R_V=3.1$). This value is similar to the interstellar extinction derived by \citet{car2022}, i.e., $A_V \approx 2.8\pm 0.4$ mag.

The reflection nebula itself appears to have some structure. At least one globule is visible within the nebula but it is not in emission (Fig. \ref{fig:nebdraw}). Their nature is unclear. A full analysis of MUSE data on the reflection nebula will be the basis of a forthcoming paper.



\section{Discussion}\label{sec:discussion}

\subsection{Accretion disk model}\label{model}

The MATISSE observations revealed a marginally resolved structure toward the center, smaller than 2.5 mas (Sect. \ref{sec:matresults}) or a diameter of $<3$~au), which we interpret as an inner accretion disk, a usual component of FUor disks \citep{hartmann1996}. Since the interpretation of these MATISSE snapshot observations is model-dependent, we opted to simulate the inner accretion disk with a typical, steady-state, geometrically-thin but optically-thick disk model as in \citet{lykou2022}. 

\begin{figure}[hbtp]
    \centering
    \includegraphics[width=\columnwidth]{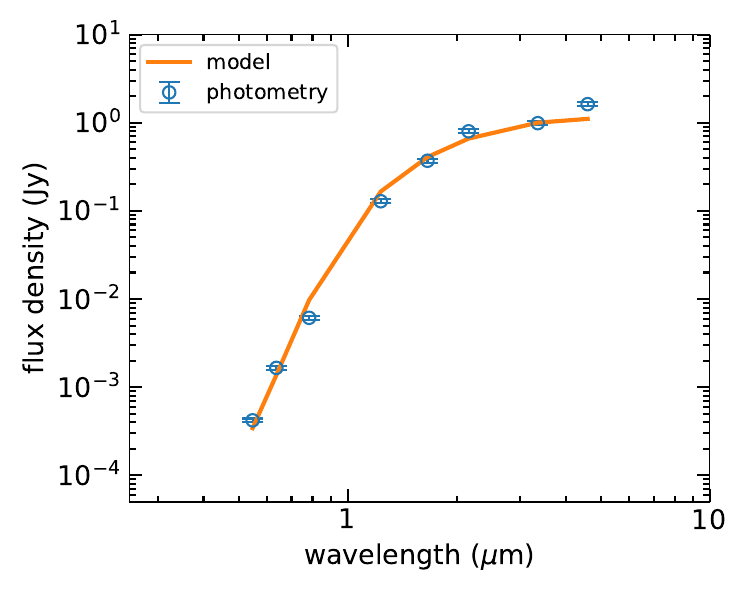}
    \caption{Accretion disk model against the 2019-2021 photometry of V900 Mon from MUSE, SMARTS, and NEOWISE.
    }
    \label{fig:modelsed}
\end{figure}

The disk model was fitted to the MUSE synthetic photometry, the SMARTS $JHK_s$ photometry (Table~\ref{tab:nirphot}), and the averaged 2019-2021 NEOWISE photometry ($W1$ and $W2$ bands) that was corrected for saturation (Fig.\ref{fig:modelsed}). While these observations are not simultaneous, they were obtained within a few years and according to the lightcurve (Fig.~\ref{fig:allphot}) V900 Mon brightness was relatively constant during that time. Therefore, we could combine these observations to an SED.

For the case of V900~Mon, the fitted parameters (Table~\ref{tab:accmodel}) were the product of the stellar mass and the mass accretion rate ($M_* \dot{M}$) and the extinction ($A_{\rm V}$). The fitted parameters' uncertainties were estimated with a Monte Carlo approach assuming 0.05 mag uncertainty on each measured photometric point, and then allowing 100 iterations on slightly modified photometric data randomly tuned within the error bars. 

The disk's inclination was fixed at 14\degr\ as estimated from the MATISSE $N$-band data. We also attempted to fit the disk at the inclination suggested by ALMA data (Table~\ref{tab:alma_sizes}) but the results were similar for such a pole-on disk. The disk outer radius, $R_{\rm out}$, was fixed at 1.5 au based on the angular size estimates in $L$-band.  Overall, extending the outer radius further does not modify the fit to the SED, since the majority of the disk flux originates within a region of 1-1.5 au.. The inner radius, $R_{\rm in}$, was fixed equal to the stellar radius at $R_* =2$~R$_\sun$. The latter is based on the FU~Ori model of \citet{lykou2022}, presuming V900~Mon is another typical FUor. However, the inner radius cannot be constrained further and this is a known degeneracy for this model\footnote{This may be alleviated from future measurements of the hot inner disk's size at shorter wavelengths (i.e., $\le 2\mu$m).}. The best-fit model from our parametric search is shown in Fig.~\ref{fig:modelsed}. By integrating the synthetic SED, we calculated an accretion disk luminosity, $L_{\rm disk}=\rm 314~L_\sun$.

A first assessment on the extinction was made using a similar comparison to \citet{connelley2018} between the FU~Ori SpeX/IRTF spectrum (Connelley, priv. comm.) and our V900~Mon spectrum from the same instrument (see Sect.~\ref{irtf}) between 1 and 4 \micron. We cannot confirm a circumstellar extinction as high at 13.5 mag as derived by \citet{connelley2018}, as we note that the slope of the V900~Mon spectrum increases toward the blue (that is below 1.4 \micron ) when $A_V > 9$ mag. 

For the given parameters and photometry, the accretion disk model predicts an extinction $A_V = 8.80\pm0.06$ mag and $M_* \dot{M} = (4.1\pm0.1)\times10^{-5}\,\rm M_\sun^2 \, yr^{-1}$. Assuming a stellar mass of 1~M$_\sun$, the accretion rate $\dot{M}$ becomes $\approx 4\times10^{-5}\,\rm M_\sun \, yr^{-1}$. If the eruption occurred 30 years ago and the accretion rate was stable at this value for this 30-year period, then the star has already accreted approximately $1\,\rm M_{jup}$.

This accretion disk model can give a first approximation of the inner disk's properties based on the current observables. Future interferometric observations (e.g., $K$-band with GRAVITY/VLTI for hydrogen and CO gas) would help constrain further the accretion disk's properties. 


\begin{table}[htbp]
    \centering
    \caption{Accretion disk model parameters. The fit presented here has $\chi^2=30.87$.}
    \begin{tabular}{lcc}
    \hline\hline
    Parameter & Value & Status \\ \hline
    $R_*$ & 2~R$_{\odot}$ & Fixed\\
    $R_{\rm in}$ & $\equiv R_*$ & ''\\
    $R_{\rm out}$ & 1.5~au & ''\\
    $i$ & 14\degr & ''\\
    $M_* \dot{M}$    & $(4.1\pm0.1)\times 10^{-5}$ M$_{\odot}^2$\,yr$^{-1}$ & Fitted\\
    $A_V$ & $8.80\pm0.06$ mag & ''\\
    \hline
    \end{tabular}
    \label{tab:accmodel}
\end{table}

\subsection{Dusty disk}\label{sec:mine}
In Sect.~\ref{sec:matresults} we showed the radial variation of the correlated spectra with respect to baseline length in the MATISSE and MIDI data. Protoplanetary disks in FUors are silicate rich, however this is not clear from the correlated spectra. In the following, we disentangle the 10 \micron\ silicate feature from the spectra and examine any variations with respect to baseline length.

\begin{figure*}
    \centering
\includegraphics[width=0.75\linewidth]{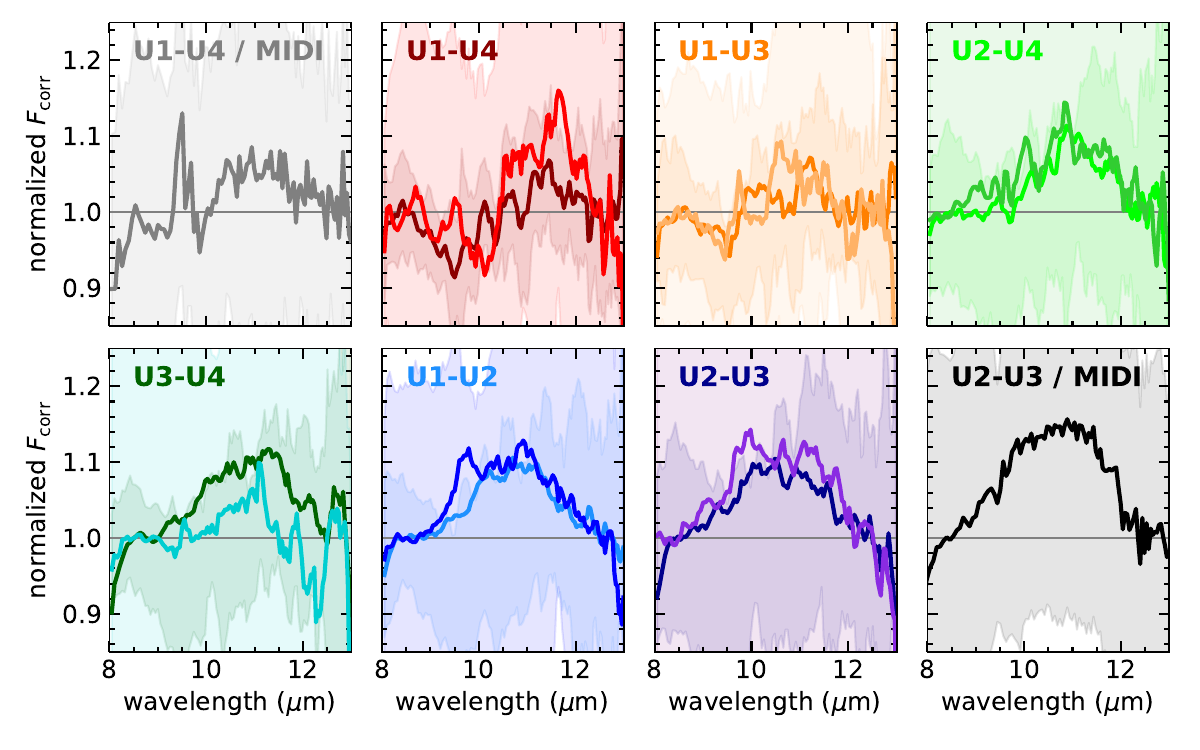}
    \caption{MATISSE and MIDI normalized continuum-subtracted correlated spectra per baseline. From top-left and clockwise the spectra are plotted from the longest baseline to the shortest, while an offset has been applied for clarity. The MATISSE 2019 data are drawn in darker shades of color compared to the respective 2020 data that are shown in lighter shades. The MIDI data are shown as black and gray lines. The uncertainties of the normalized correlated fluxes are shown for comparison as shaded regions.
    }
    \label{fig:norm}
\end{figure*}

\subsubsection{Radial distribution of amorphous silicates}\label{silemi}

We followed the method of \citet{vanboekel2003} to enhance the 10 \micron\ silicate feature by subtracting the continuum emission from each baseline. The continuum was linearly fit between 8.5 and 12.5 \micron\ at each baseline. This allows the removal of the more noisy part of the spectrum at longer wavelengths, an effect that appears to be common in MATISSE observations of faint targets (i.e., $\leq5$~Jy). The continuum is subsequently subtracted from the correlated spectrum, and then the correlated spectrum is normalized by dividing it with the mean value of the continuum. These normalized and continuum-subtracted correlated spectra are shown in Fig.~\ref{fig:norm}. Here, the spectra have been plotted from top to bottom and left to right in decreasing baseline length -- from U1-U4 ($\sim$130 m) to U2-U3 ($\sim$40 m) -- and an offset has been applied for clarity. Darker colors correspond to the 2019 data, while lighter colors to the 2020 data. Also plotted for reference are the MIDI spectra (black and gray lines). The uncertainties of the correlated spectra are shown as `shaded' regions in Fig.~\ref{fig:norm} indicating the low S/N for V900 Mon.  

A change of the strength of the 10 \micron\ silicate emission feature with respect to each baseline is noticeable in these normalized spectra (Fig.~\ref{fig:norm}). The feature is stronger and in emission at the largest spatial scales of the disk (i.e., $r<18$ au) as indicated by the shortest baselines (U2-U3 and U1-U2). It presents the typical trapezoidal shape expected for larger-sized grains (grain size $\geq 1$ \micron) \citep[e.g.,][]{bouwman2003}. 

The strength of the feature diminishes at smaller spatial scales and at the intermediate-length baselines (U3-U4 and U2-U4), while it appears to be near-zero or even negative at the smallest spatial scales (i.e., $r<7$~au) and at the maximum baseline length (U1-U2 and U1-U4). This variation is supported by the archival MIDI correlated spectra for which a similar process was followed (also in Fig.~\ref{fig:norm}). The short-baseline spectrum (U2-U3) also indicates a trapezoidal shape as in the MATISSE data, while the long-baseline spectrum (U1-U4) are more heavily affected by the terrestrial ozone layer (9.4 -- 9.9 \micron). It is worth noting that the peak of the normalized spectrum for the shortest baseline is near 11.4 \micron, unlike the MATISSE spectrum. However, the MIDI spectrum suffered from an abrupt jump in flux between 11.7 and 12.2 \micron, which could have affected the final shape of the normalized spectrum.

Considering that the disk of V900~Mon is viewed nearly pole-on, this suggests that silicate emission from larger grains originates from the disk surface at radii larger than 10~au from the star, while at the inner-most regions of the disk ($r<7$~au), the silicate emission is either absent or this region is self-shielded by optically thicker material consisting of small-sized grains. 

\begin{figure}
    \centering
\includegraphics[width=\columnwidth]{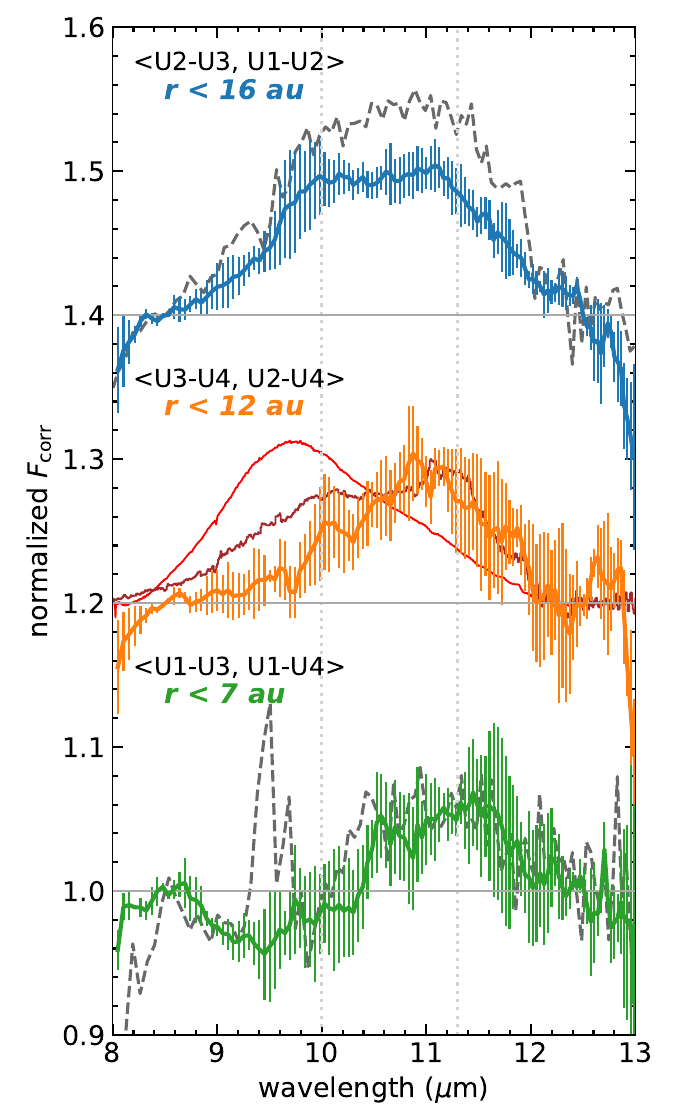}
    \caption{MATISSE normalized and averaged spectra (both epochs) of Fig.~\ref{fig:norm} per different baselines as indicated in the legend. From top to bottom, the three spectra (blue, orange, and green) correspond to different regions of the protoplanetary disk within radii of 16, 12, and 7 au, respectively. An offset has been applied for clarity. Over-plotted for comparison are the MIDI normalized spectra (gray; dashed), the profile of ISM silicates \citep[red line, ][]{kemper2004}, and the normalized spectrum of comet Hale-Bopp (brown line) . The two dotted vertical lines act as simple guides for the eye at 10 and 11.3 \micron.
    }
    \label{fig:MATave}
\end{figure}

This radial variation can be elucidated further when averaging baselines (from Fig.~\ref{fig:norm}) that cover similar spatial scales -- and therefore different disk regions -- over both epochs. These averages are shown in Fig.~\ref{fig:MATave}. In more detail, these show the average of the shortest baselines (U2-U3 and U1-U2) that probe regions $r<16$~au (blue line), then intermediate baselines (U3-U4 and U2-U4) which probe regions $r<12$ au (orange line), and finally the longest baselines (U1-U3 and U1-U4) for $r<7$ au (green line). Furthermore, we compared these to the MIDI normalized correlated spectra (gray dashed lines) at similar baselines (Fig.~\ref{fig:norm}). The silicate feature not only weakens, but its shape changes as well: the 9.7 \micron\ part weakens, while the longer wavelength part remains stable, and the overall shape becomes more triangular with a peak around 11 \micron. The error bars shown in Fig.~\ref{fig:MATave} are the standard deviations of the averaged spectra. Further mineralogical analysis follows in the next section.

\subsubsection{A lack of crystalline silicates}\label{cryst}

Crystalline silicates appear to be common in protoplanetary disks around Herbig and T~Tauri stars \citep[e.g.,][]{vanboekel2005,olofsson2009} but are apparently absent from disks around FUors \citep[e.g.,][]{quanz2007,kospal2020}. However, crystalline silicates have been found in other types of eruptive stars  \citep[e.g., EX Lupi, ][]{abraham2009}.

For the case of V900~Mon, \citet{varga2018} analyzed the uncorrelated MIDI spectra from 2013, which represent silicate emission inside and outside a radius of 18~au, but their results were inconclusive due to low S/N. \citet{kospal2020} analyzed VISIR/VLT mid-infrared spectra that represent silicate emission over the entire disk, but concluded that V900~Mon, like all FUors, contains large amorphous grains.

We examined the MATISSE normalized spectra of Fig.~\ref{fig:MATave} for any signs of crystalline silicate emission by over-plotting the spectrum of comet Hale Bopp (brown line) that contains high fraction of crystalline silicates (e.g., forsterite). For comparison, we also show the Galactic Center spectrum of small-sized amorphous silicate grains \citep[red line,][]{kemper2004}. There is no noticeable emission at the typical narrow emission band at 11.3 \micron\ in any of the baselines at this S/N levels.  Overall, we see no unequivocal signs of crystalline silicates in the MATISSE spectra.


\subsubsection{Circumstellar extinction}\label{g21quanz}

Our accretion disk model suggested a circumstellar extinction of $A_V\sim 9$ mag (Sect.~\ref{model}). However, the silicate feature is clearly seen in emission at the largest spatial scales (Fig.~\ref{fig:MATave}) and within a FOV of 0.5\arcsec , which contradicts such high extinctions. 

\citet{quanz2007} noted that the silicate profiles of FUors that are in emission, once corrected for interstellar extinction, have a similar shape. Here, we built on this result and assume that indeed the spectral shapes are identical. As shown earlier, the silicate feature disappears below 10.5 \micron\ at the smallest spatial scales. We explored whether varying amounts of extinction can account for the variation of the silicate feature at different spatial scales compared to our baseline feature, that is the typical trapezoidal shape shown in blue in Fig.~\ref{fig:MATave}. We opted for the extinction law of \citet{gordon2021} which covers the mid-infrared regime.

\begin{figure}
    \centering
    \includegraphics[width=\columnwidth]{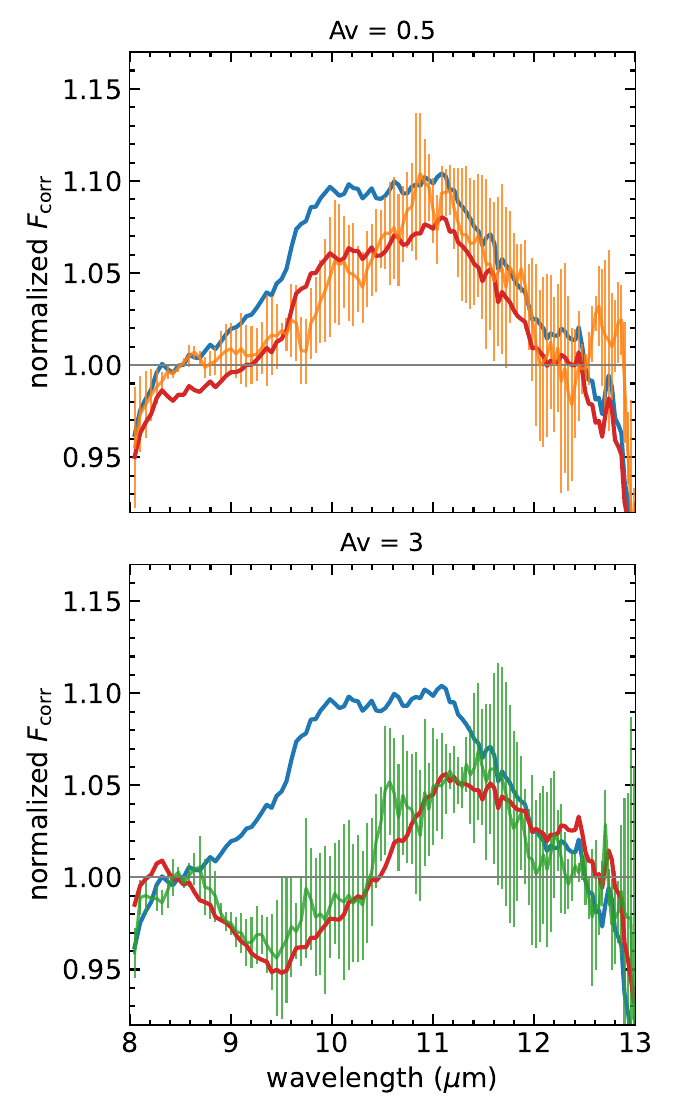}
    \caption{The averaged and normalized correlated spectra (blue) of the shortest baselines (cf. top panel in Fig.~\ref{fig:MATave}) reddened with the extinction law of \citet{gordon2021} and $A_V=0.5$ and 3 mag in the top and bottom panels, respectively. The reddened spectra (red) are a good fit to the averaged-normalized spectra from the intermediate (orange; top panel) and longest baselines (green; bottom panel). }
    \label{fig:g21ext}
\end{figure}

Since the disk is seen pole-on, let us presume that the silicate feature should be in emission over its entire surface. Hence, we reddened the averaged spectrum (seen in blue in Fig.~\ref{fig:MATave}) by different $A_V$ with the extinction law described above. As it is shown in Fig.~\ref{fig:g21ext}, we can match\footnote{Although the fitting was performed `by eye' and no optimization was made, it was found to be sufficient in matching the average levels.} the averaged and normalized correlated spectra for the two other baselines (those shown in orange and green colors in Fig.~\ref{fig:MATave}) by applying $A_V = 0.5$ and 3.0 mag, respectively. If we presume that the interstellar extinction is 3.4 mag (Sect.~\ref{knot} and \ref{model}), then we find that the total extinction in the line of sight is approximately $A_V = 6.4$ mag toward the central 10 au radius area of the star. This demonstrates that the extinction varies depending on the line of sight. We explore this in the following sections.


\subsection{A hidden companion, a stratified disk, or a dust clump?}\label{dustclump}

\subsubsection{Companion}
An alternative scenario besides the localized extinction to explain the spectral shapes at the longest baselines would be the presence of a companion. A companion with sufficient separation and flux ratio might be responsible for the modulation-like signal in the long-baseline MATISSE normalized correlated spectra (bottom panel in Fig. \ref{fig:MATave}). Theoretical works suggest that FUor outbursts can be triggered by flyby events, where a companion transverses through the protostellar disk triggering increased accretion onto the protostar \citep{bonnell}. Signatures of such flyby stars are streams of material seen in scattered light, such as in the case of Z~CMa \citep{canovas2015} for which \citet{dong2022} identified a strong candidate source as a flyby star. At present, we are not aware of any polarimetric imaging observations of V900~Mon that could potentially reveal such features. 

If we presume a flyby star traveling at 10~km~s$^{-1}$ might have interacted with V900~Mon in the last century, then this star could be found within a maximum distance of 175~mas (210~au) from the star. Such a distance is potentially within the field-of-view of the UTs with MATISSE ($0.5$\arcsec\ at 10 \micron). However, that is also within the capabilities of ALMA although neither \citet{takami2019} nor \citet{kospal2021} found any wide-orbit companion, that is one outside V900 Mon's disk (i.e., beyond $\ge50$ mas based on its deconvolved size). If a flyby companion exists within 50 mas from the protostar, then this ought to be bright in both $L$ and $N$ bands to be detectable in the mid-infrared, normalized correlated spectra, as well as at a distance larger than at least one resolution element in the $N$-band (i.e., $\approx10$ mas).

Previous studies have shown that MATISSE is able to detect companions at a flux ratio as low as 2\% within an orbital separation of 100~mas in $L$-band \citep{matisse}, however those estimates resulted from higher S/N data. The lack of a distinct sinusoidal modulation in the $L$- and $N$-band closure phase signals (Figs.~\ref{fig:Lcps} and \ref{fig:Ncps}), suggests that no companion was detected with these MATISSE observations. Attempts to estimate the detectability of a companion with the current data set using the methods of \cite{candid} and \cite{pmoired} have been inconclusive due to (a) the limited uv-coverage and therefore the expected degeneracy of any solution regarding the location of a companion, and (b) the low brightness of the science target and the known instrumental biases for MATISSE. Taking into account these degeneracies, a potential companion ought to be much brighter than $ \gg 5\%$ to be detectable in this data but no such signature is seen here. Perhaps future observations at higher S/N may help confirm or disprove the presence of a flyby star within 10--50 mas from V900 Mon. For such a faint target, MATISSE/GRA4MAT observations would be recommended.

\subsubsection{Previous temperature gradient models }

\citet{varga2018} modeled the 2013 MIDI short-baseline data with a thin, flat, passive disk of fixed outer radius (300~au), dust sublimation temperature at 1500~K, and a power-law temperature distribution ($T\propto R^{-0.69}$). They calculated a dust sublimation radius $\approx 1.4$~au, which translates to about 1.3~mas for their adopted distance. This radius is of similar size to the radius of the disk region emitting in the $L$-band, as derived by centro-symmetric brightness distribution fitting of the MATISSE data, and thus comparable to the size of the simulated accretion disk (Sect.~\ref{model}).

Their disk model indicates that half of the mid-infrared flux is emitted within a radius of $6.4^{+3.0}_{-1.8}$~au. However, this would be roughly the radius at which the silicate feature is in absorption (Fig.~\ref{fig:MATave}). Consequently, either 
the majority of the emission arises from its edge, or the simple temperature-gradient model (although it can fit the MIDI data) is not sufficient.


\subsubsection{Dust clump}

\begin{figure}[btp]
    \centering
    \includegraphics[width=\columnwidth]{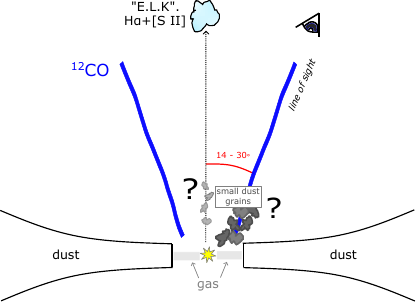}
    \caption{Cartoon indicating the presumed geometry of the system.  The drawing is not to scale.}
    \label{fig:cartoon}
\end{figure}

In Section \ref{silemi}, we showed that there may be material shielding the innermost regions of the protoplanetary disk. Since the disk is oriented nearly pole-on, this material could be interpreted as a ``clump'' of small dust grains on top of the protoplanetary disk, or else at the origin of the molecular outflow emanating from the disk \citep{takami2019}. We therefore presume that this material is in our line-of-sight. A sketch of this assumed geometry is shown in Fig.~\ref{fig:cartoon}. A similar geometry has been postulated for the Herbig Ae star HD163296, which has Herbig-Haro objects and where the star presumably suffered a dimming event due to a dust cloud \citep{ellerbroek2014, pikhartova2021}.

Although the exact structure and composition of these silicate grains cannot be constrained from the MATISSE spectra, based on  our analysis in Section \ref{silemi}, we assume that this material is primarily composed of small and spherical amorphous silicate dust grains with a grain size $a \leq 0.1$ \micron. If the circumstellar extinction is about $A_V=3$ mag (Sect.~\ref{g21quanz}), then following the \citet{gordon2021} law, we calculate an extinction of $A_{\rm sil} \approx 0.25$ mag for the 10 \micron\ silicate feature. The optical depth of the silicate feature is $\tau_{\rm sil} \approx 0.23$.

We estimate the dust mass using $\tau_{\rm sil} = \rho_{\rm dust}\, \kappa_{\rm abs}\, \ell$, where $\rho_{\rm dust}$ is the dust mass density and is equal to $M_{\rm dust}\, V^{-1}$ for a dust mass $M_{\rm dust}$ in a volume $V$, $\ell$ is length along the line of sight, and for small-sized ($a \leq 0.1$ \micron) amorphous silicate dust, the mass absorption coefficient\footnote{Value derived with the {\tt OpacityTool} \citep[cf.][and references therein]{woitke2016}.} $\kappa_{\rm abs}$ is $\approx 4000$\,cm$^{2}$\,g$^{-1}$. It is easy to show that for a cylinder of radius $r$ and its volume calculated along the line of sight, the above equation becomes $M_{\rm dust} = \pi r^2 \, \tau_{\rm sil}\, \kappa_{\rm abs}^{-1}$. Since the material is constrained within a radius $r\leq5$~au, the amount of dust enclosed becomes $M_{\rm dust}\approx 9.9\times 10^{23}$~g. This is approximately the mass of the dwarf planet Ceres. For a higher circumstellar extinction (Sect.~\ref{model}), the dust mass is  slightly larger at $\approx 2.9\times 10^{24}$~g. This is a tentative estimate since: (a) we assumed a mass absorption coefficient for amorphous silicate grains of a certain size, (b) we did not take into account the disk's inclination\footnote{A slightly higher inclination at 30\degr\ would reduce this mass estimate by about 14\%.} and the fact that only part of the disk might be obscured, (c) the dust volume could be much smaller than what is assumed above, and (d) the disk morphology and origin of the CO outflow may be more complex than what is assumed in this simplistic analysis. We expect that follow-up imaging in the near-infrared and sub-millimeter wavelengths may shed more light into the complexity of the disk and its surrounding region, followed by future explorations of radiative transfer simulations of the protoplanetary disk's properties.

\subsection{Collimated outflow: traces of a jet?}\label{sec:jet}

We analyzed archival images of V900~Mon from VPHAS+ in $ugri$ and H$\alpha$ filters, from the Panoramic Survey Telescope and Rapid Response System \citep[Pan-STARRS1; ][]{ps1} in $grizy$ filters, and UKIDSS in $JHK$ filters. Although the surveys differed in terms of sensitivity, filters, and telescope aperture used, we identified common features overall, the most prominent being a ``helicoidal'' tail that fans out beyond 20\arcsec\ west-southwest from the star and that is seen at all wavelength ranges although it is brighter in the visual. Part of this tail is also seen by MUSE, although the chosen FOV is smaller than the full extent of the nebula to the west\footnote{See for example the deep-exposure images by amateur astronomer Adam Block, \url{https://skycenter.arizona.edu/astrophotography/lbn1022}}
    
In Section~\ref{sec:muse}, we showed the emission-line knot discovered in the MUSE data. \citet{reipurth2012} obtained [\ion{Fe}{ii}] and H$_2$ wide-field images with the Gemini NIRI camera and a wider FOV (120\arcsec$\times$120\arcsec) compared to that of MUSE, however they did not see any typical jet-like features near V900~Mon. We offer two scenarios for this non-detection. Either (a) the knot does not emit at those wavelengths, or (b) its surface brightness was too low to be detected. For the first case, the knot's density might not be high enough, since [\ion{Fe}{ii}] traces denser gas ($n_e \sim 10^4$~cm$^{-3}$). For the latter case, a higher-sensitivity and higher-spectral resolution near-IR spectrum (e.g., ERIS/VLT) of the knot is required to answer these questions. Conversely, none of the SpeX/IRTF spectra covered that region.

Coincidentally the knot is aligned with one of the globules seen in the white light and H$\alpha$ MUSE images (Figs.~\ref{fig:muse_cont_im} and \ref{fig:nebdraw}) at a PA of 250\degr\ ($\pm5$\degr) east-of-north from the stellar position. This is nearly perpendicular to the PA of the disk's major axis (see also Table~\ref{tab:alma_sizes} and Fig.~\ref{fig:cartoon}). However, the globule might be a localized enhancement of scattered light emission in the nebula unrelated to the emission-line knot. \cite{takami2019} shows that the CO (2-1) gas emission indicates two wide cavities, where the blue-shifted one has an opening angle of 70\degr. The angular resolution of those maps is superior to that of MUSE, while the FOV is also smaller (12\arcsec\ against 1\arcmin). Figure~\ref{fig:linemap} shows that the knot detected by MUSE is aligned to the CO (2-1) emission from \citet{takami2019}.

We searched for any additional knots that may have been produced at the last eruption 30 years ago. Assuming that such a knot has an average velocity of 25~km\,s$^{-1}$ -- similar to the knot discovered by MUSE -- it would have moved by about 0.13\arcsec. That distance is smaller than a single MUSE WFM spaxel and as such very close to the eruptive star. Therefore, these MUSE data are inconclusive but future observations in the Narrow Field Mode of MUSE could potentially shed more light.

There are other examples of FUors with jet-like emission associated with distant knots. \citet{andreasyan2021} suggest that V565~Mon is another eruptive star associated with Herbig-Haro globules that were first identified near its reflection nebula, Parsamian 17 \citep{magakian2008}, and these are also identifiable in the Pan-STARRS1 $gri$ images. Its stellar spectrum is similar to V900~Mon. However, V900~Mon does not appear to have globules that far from the star based on the VPHAS+ and Pan-STARRS1 pseudo-color maps, but that could be due to low S/N in those surveys. Perhaps the most prominent example is Z~CMa, a binary system composed of two young eruptive stars one of which the South-East component is an FUor. This FUor launched a micro-jet with an average velocity of $\sim200$\,km\,s$^{-1}$ seen in the near-infrared [\ion{Fe}{ii}] 1.64 \micron\ \citep{whelan2010}, that itself is aligned with [\ion{S}{ii}] emitting knots located about 60\arcsec\ from the star \citep{poetzel1989} alleging to earlier eruptions from that system.

Based on the kinematic age of the knot (Sect.~\ref{sec:muse}), we can deduce at least one earlier eruption just over 1000 years ago. The mechanism that created such a knot is ambivalent, which obviously cannot be easily attributed to a previous outburst.

The [\ion{S}{ii}] emission alleges to shock excitation and the co-alignment of the fast-moving gas ($\sim100$~km~s$^{-1}$) with a much slower \citep[$< 20$~km~s$^{-1}$, ][]{takami2019} and relatively wider molecular outflow suggests the presence of a jet inside said outflow. If indeed these components constitute jet-like emission, then this would imply that the jet was collimated by a magnetic field at the central engine, that is near the star and above the disk \citep[cf. the review of ][ and references therein]{frankPP6}. Although jets are more common at other stages of protostellar evolution, there are at least four examples of jets in FUors (Z~CMa,  V346 Nor, SVS 13A, L1551 IRS5). Nevertheless, it is thought that FUors are not magnetically active during eruptions, since the magnetic field lines are suppressed by excess ram pressure from the material accreted from the disk onto the protostar \citep[cf., ][ and references therein]{hartmann2016}. Our current observations cannot provide a concrete solution to this contradiction.


\section{Conclusions}\label{sec:concl}
We reported on our recent observations of the eruptive star V900~Mon with MATISSE/VLTI and MUSE/VLT. The MATISSE $L$-band observations suggest a marginally resolved source with an angular size $<3$ mas at 3.5 \micron , providing an upper limit of 2~au for the radius of the disk region emitting at this wavelength. The region of the passive dusty disk that emits in the $N$-band is confined within a radius $<20$ au from the star. Geometric model fits to the MATISSE $N$-band data corroborate the disk's pole-on geometry previously found by ALMA.

The interferometric observations with MATISSE, which are supported by archival MIDI data, revealed the radial variation of the 10 \micron\ silicate feature. Similar to earlier works, we could not find signatures of crystalline silicates in the disk. The silicate feature is clearly seen in emission at large spatial scales (disk radius $\geq10$ au) with a trapezoidal profile indicative of amorphous, large-sized dust grains ($a\geq 1$ \micron). The shape of the silicate feature's spectral profile changes toward the inner regions of the disk, and it diminishes below 10.5 \micron\ at spatial scales $\leq5$~au. Our analysis of this spectral signature suggests that the innermost region of the disk is shielded by higher extinction, possibly by an accumulation of small dust grains (size $\leq 0.1$ \micron ) that may be located near or covering a portion of the inner disk.  Taking this into account, the small grains may be located at or near the origin of the collimated CO outflow found by ALMA.

Furthermore, we report the discovery of an emission-line knot from our MUSE observations. Its size is approximately 1.30\arcsec$\times$0.98\arcsec . This knot is co-aligned to the collimated CO outflow previously found by ALMA (PA $=250\pm5$ deg, east-of-north), and its projected separation from the star is approximately 27,000~au or 0.13~pc. The most prominent lines are H$\alpha$, [\ion{N}{ii}] at 6584\AA, and the [\ion{S}{ii}] 6716/30\AA\ doublet. There is a tentative detection of [\ion{O}{i}] at 6300\AA\, but it appears heavily blended with terrestrial emission. All emission lines are blue-shifted with respect to rest wavelengths with an average velocity of 100~km~s$^{-1}$. The knot's kinematic age is approximately 5150 years. Although this is an upper limit, it alludes to an earlier eruption of V900~Mon compared to its current one 30 years ago. The presence of [\ion{O}{i}] and [\ion{S}{ii}] emission suggests that the knot is shock excited, indicating that the FUor drives a jet. However, no jet-tracing emission lines were detected in the vicinity of the star suggesting that no such activity occurred during or immediately before the current eruption 30 years ago. If the knot is shock-excited from a mechanism originating from V900 Mon's disk, and since the disk itself is seen nearly pole-on, then the dusty ``shield'' found by MATISSE ought to be off-centered from the star.

V900~Mon is  still in eruption and is therefore an active laboratory for the exploration of FUor evolution. All of the above suggest that further investigation of this eruptive star is necessary to assess the mechanism that shields the inner parts of its accretion disk and what mechanism launched the knot more than a millennium ago. For the former, we suggest future observations of scattered-light imaging with SPHERE/VLT, and interferometric imaging of the CO and hydrogen gases in the accretion disk with GRAVITY/VLTI. At the moment the MUSE WFM observations cannot disentangle minute kinematics and proper motion within the knot, however, a study of the knot at higher spatial and spectral resolution is possible with MUSE NFM mode and/or XSHOOTER/VLT in the visual, as well as the integral field unit of ERIS/VLT in the near-infrared.

\begin{acknowledgements}

This project received funding from the Hungarian NKFIH OTKA project no. K-132406, and from the European Research Council (ERC) under the European Union's Horizon 2020 research and innovation programme under grant agreement No 716155 (SACCRED). FCSM received financial support from the European Research Council (ERC) under the European Union’s Horizon 2020 research and innovation programme (ERC Starting Grant "Chemtrip", grant agreement No 949278). This work is supported at The Aerospace Corporation by the Independent Research and Development Program. M.S and M.P. were supported by NASA grant NNX16AJ75G. J.V. is supported by NOVA, the Netherlands Research School for Astronomy. S.K. acknowledges support from ERC Consolidator Grant "GAIA-BIFROST" (Grant Agreement ID 101003096) and STFC Consolidated Grant (ST/V000721/1).

MATISSE was designed, funded and built in close collaboration with ESO, by a consortium composed of institutes in France (J.-L. Lagrange Laboratory – INSU-CNRS – C\^ote d’Azur Observatory – University of C\^ote d’Azur), Germany (MPIA, MPIfR and University of Kiel), the Netherlands (NOVA and University of Leiden), and Austria (University of Vienna). The Konkoly Observatory and Cologne University have also provided some support in the manufacture of the instrument. This research has made use of the services of the ESO Science Archive Facility.

\end{acknowledgements}
%
\bibliographystyle{aa} 
\bibliography{v900.bib} 
%


\begin{appendix} 

\section{Tables}

In this appendix we present the VLTI observing logs for the MATISSE and the archival MIDI data (Table~\ref{tab:logvlti}) and for our optical and infrared observations (Table~\ref{tab:logphot}). The evolution of V900~Mon in the last 30 years was shown in Sect.~\ref{sec:evol}. Our near-infrared photometry is presented in Table~\ref{tab:nirphot}.


\begin{table}[hbtp]
    \centering
    \caption{Near-infrared photometry}
    \begin{tabular}{cccc}
    \hline\hline
    Julian date & Band & $m$ (mag) & error \\ \hline
    \multicolumn{4}{c}{TCS} \\
2455989.387 & $J$ & 10.419 & 0.008 \\ 
2456210.750 & $J$ & 10.501 & 0.016 \\ 
2456211.758 & $J$ & 10.441 & 0.013 \\
2456212.746 & $J$ & 10.454 & 0.007 \\
2456213.762 & $J$ & 10.465 & 0.017 \\ 
2456214.738 & $J$ & 10.449 & 0.016 \\ 
2458432.727 & $J$ & 10.357 & 0.017 \\ 
2455989.391 & $H$ &  8.814 & 0.006 \\ 
2456210.754 & $H$ &  8.818 & 0.007 \\ 
2456211.766 & $H$ &  8.793 & 0.003 \\ 
2456212.750 & $H$ &  8.791 & 0.013 \\ 
2456214.742 & $H$ &  8.854 & 0.015 \\ 
2458432.730 & $H$ &  8.663 & 0.045 \\ 
2455989.398 & $K_s$ &  7.756 & 0.024 \\ 
2456210.762 & $K_s$ &  7.820 & 0.028 \\ 
2456211.770 & $K_s$ &  7.653 & 0.030 \\ 
2456212.758 & $K_s$ &  7.723 & 0.020 \\ 
2456214.750 & $K_s$ &  7.713 & 0.012 \\ 
2458432.742 & $K_s$ &  7.522 & 0.017 \\ \hline
    \multicolumn{4}{c}{REM} \\
2457609.902 & $J$ & 10.114 & 0.063 \\ 
2457609.906 & $H$ &  8.527 & 0.017 \\ 
2457609.410 & $K_s$ &  7.387 & 0.034 \\ \hline
    \multicolumn{4}{c}{IRTF$\dagger$} \\
2457409.5 & $K$ & 7.35 & 0.04 \\
2457410.5 & $K$ & 7.34 & 0.04 \\
2457412.5 & $K$ & 7.41 & 0.04 \\
2457413.5 & $K$ & 7.34 & 0.04 \\ \hline
    \multicolumn{4}{c}{SMARTS} \\
2458495.564 & $J$ & 10.233 & 0.018 \\ 
2458495.669 & $H$ &  8.604 & 0.037 \\ 
2458495.675 & $K_s$ &  7.303 & 0.060 \\ 
2458499.759 & $K_s$ &  7.197 & 0.056 \\ 
    \hline
    \end{tabular}
    \tablefoot{2MASS system except for ($\dagger$) MKO system. }
    \label{tab:nirphot}
\end{table}
\FloatBarrier

\begin{table*}[btp]
\caption{VLTI observing log.}\label{tab:logvlti}
\centering
\begin{tabular}{ccccccccc}
\hline
Instrument & Date & Band & $R$ & DIT & Configuration & Seeing & $\tau_0$ & Calibrator\\
& & & ($\lambda/\Delta\lambda$) & (ms)  & array & (\arcsec) & (ms) & \\ 
 \hline
\multicolumn{9}{c}{this work}\\
MATISSE & 2019-12-11T06:46:24 & L & 30 & 75 & UT  & 0.8-1.3 & 5.12 & HD59381  \\
 & " & N  & 30 & 20  & UT  & " & 5.12 & HD47667  \\
MATISSE & 2020-01-09T01:24:18 & L & 30 & 75 & UT  & 0.69-0.90 & 4.87 & HD59381 \\
 & " & N  & 30 & 20  & UT  & " & 3.72 & HD47667 \\ \hline
 \multicolumn{9}{c}{archival data}\\
MIDI & 2013-12-21T06:09:53 & N & 30 & 64 & UT2-UT3 & 0.8-1.1 & 9.53 & HD44951, HD92682 \\
MIDI & 2015-01-07T06:27:26 & N & 30 & 64 & UT1-UT4 & 0.9-1.1 & $\leq3$ & HD36673, HD44951 \\
\hline \hline
\end{tabular}
\end{table*}

\begin{table*}[btp]
    \centering
    \caption{Observing log for the optical and infrared photometry and spectroscopy.}\label{tab:logphot}
    \begin{tabular}{lcccc}
    \hline\hline
    Date &  Instrument & Band & Mode & Resolution \\ \hline
    2012-03-02 & Cain3/TCS & $JHK$ & photometry & 1\arcsec/pix \\
    2012-10-09/13 & \ldots & \ldots & \ldots & \ldots \\
    2015-01-14 & BASS/IRTF & 2.9-13.5\micron\ & spectrophometry & 0.25\arcsec -- 1.1\arcsec\ \\
    2015-01-15 & SpeX/IRTF & 0.8-5.4\micron\ & spectroscopy & 0.8\arcsec\ slit \\
    2016-01-25/26 & \ldots & \ldots & \ldots & \ldots \\
    2016-01-22/26 & Guidedog/SpeX/IRTF & $K$ & photometry & 0.12\arcsec/pix \\
    2016-08-08 & REMIR/REM & $JHK$ & photometry & 1.23\arcsec/pix \\
    2018-11-09 & Cain3/TCS & $JHK$ & photometry & 1\arcsec/pix \\
    2019-01-11/15 & Andicam/SMARTS & $JHK$ & photometry & 0.27\arcsec/pix \\
    2021-01-24 & MUSE/VLT & 0.47-0.93\micron\ & IFU & 0.2\arcsec/spaxel \\
    \hline
    \end{tabular}
\end{table*}

\section{Figures}

\begin{figure}[h!btp]
\centering
    \includegraphics[width=0.9\columnwidth]{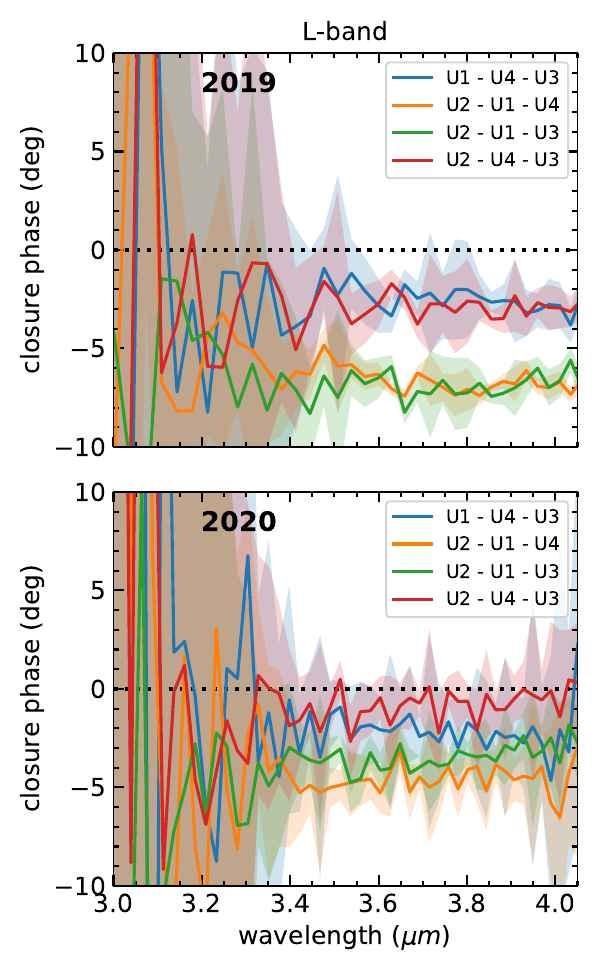}
    \caption{MATISSE closure phases in the $L$-band for both epochs. The 2019 data were affected by atmospheric variations during the observing run.}
    \label{fig:Lcps}
\end{figure}

\begin{figure}[h!btp]
        \includegraphics[width=0.9\columnwidth]{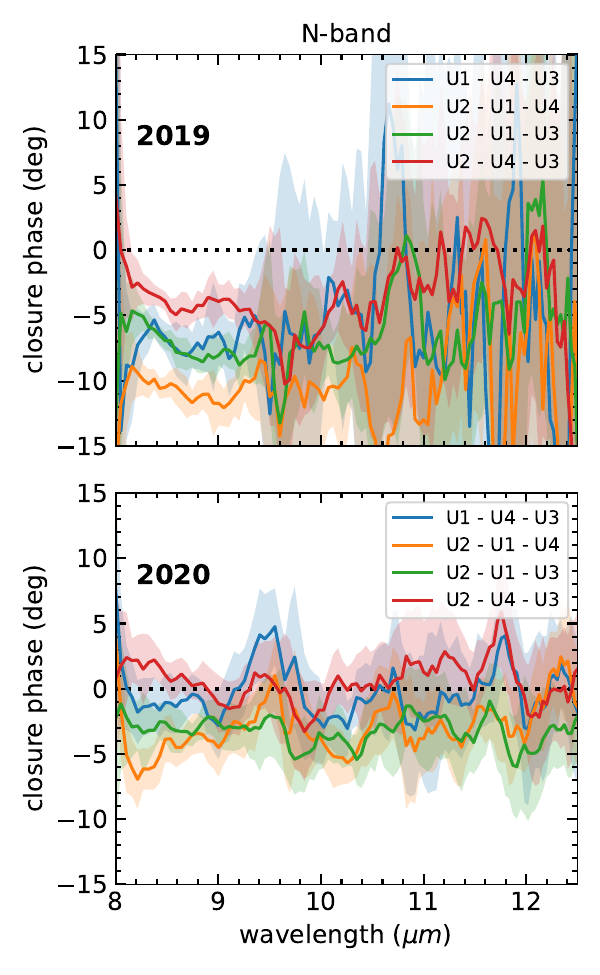}
    \caption{As in Fig.~\ref{fig:Lcps} but for the $N$-band data.}
    \label{fig:Ncps}
\end{figure}

\begin{figure}[htbp]
    \centering
    \includegraphics[width=\columnwidth]{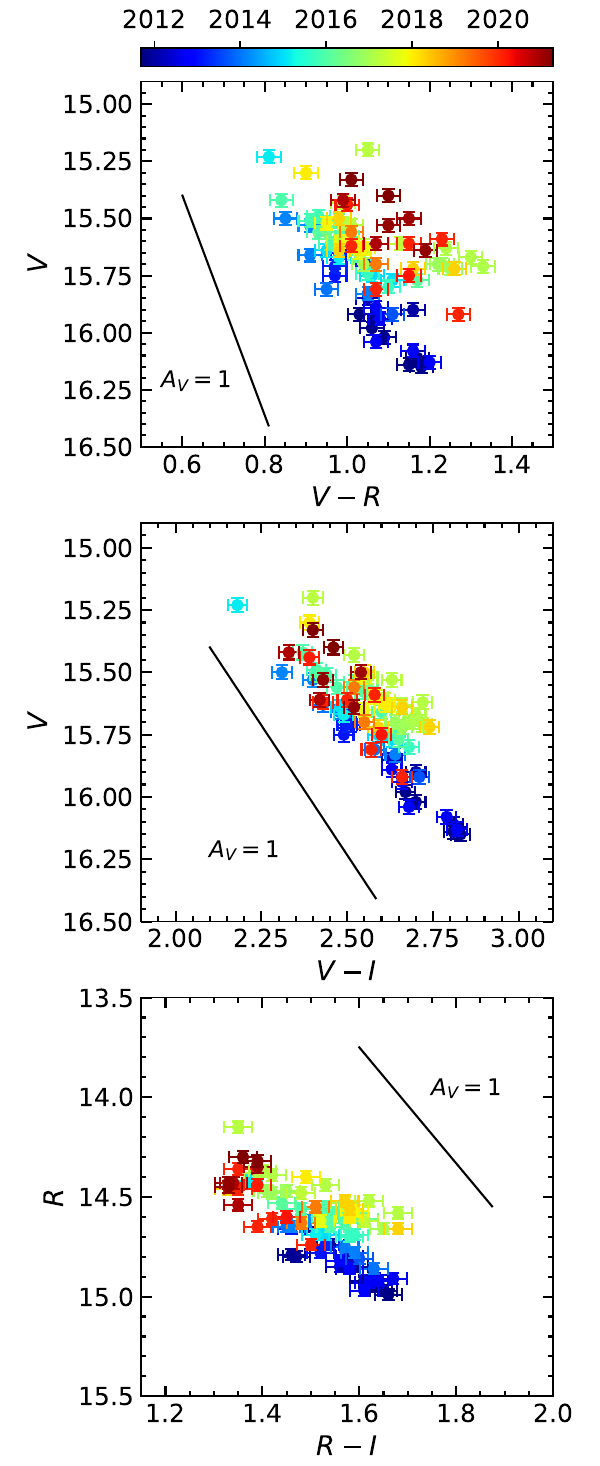}
    \caption{Color-magnitude diagrams based on the photometry of \citet{semkov2021}. The extinction vectors indicate $A_V=1$ mag based on \citet[][$R_V=3.1$]{ccm89}. }
    \label{fig:cmdiag}
\end{figure}

Supplementary MATISSE closure phases are shown in Figures~\ref{fig:Lcps} and \ref{fig:Ncps} for both epochs. The 2019 data were affected by atmospheric variations during the observing run, while the uncertainties are predominantly in excess of 1 degree. It is also worth mentioning, that in the current MATISSE pipeline the $N$-band closure phases are flipped by $\pi$ compared to the $L$-band ones due to a sign convention error \citep[see also][]{2022Natur.602..403G}. Although this does not affect the current work, since the $N$-band closure phases are not utilized, we have corrected for this flip here. Figure~\ref{fig:cmdiag} shows the color-magnitude diagrams (cf. Sect.~\ref{sec:evol}) in $V$ vs $V-R$, $V$ vs. $V-I$, and $R$ vs. $R-I$ (from top to bottom panel respectively) based on the original data of \citet{semkov2021}.

\FloatBarrier

\FloatBarrier

\section{The reflection nebula}\label{nebmorph}

The reflection nebula of V900~Mon was first pointed out in 2009 by an amateur astronomer \citet{thommes2011}. \citet{reipurth2012} indicated that emission from the nebula was barely visible in $R$-band in the first known photographic plate from the Palomar Sky Survey in 1953. Neither the nebula nor any nearby sources within 30\arcsec\ are detected in the digitized\footnote{Data for this regions are not yet available from the DASCH project.} plates from the Heidelberg galactic survey (1899-1990) and from the Bamberg Southern Sky Patrol (1955-1974). \citet{semkov2021} re-analyzed the archival, digitized plates of the POSS and UKST surveys and found that the nebula was fainter in $R$-band on 8 January 1989, however the photographic emulsions and filters differed between the two epochs. 

A careful examination of the later SuperCosmos H$\alpha$ Survey \citep[SHS; ][]{shs} indicates that the reflection nebula was already visible a decade earlier in 1999 (short red, $sr$, filter) and 2002 (H$\alpha$ filter). This would suggest that the eruption occurred sometime after 1990, since earlier data (as above) show a rather compact source at the location of the current nebula. We therefore presume that V900~Mon has been slowly erupting in the last 30 years.

\begin{figure*}
\centering
	\begin{minipage}[c]{8cm}
		\includegraphics[width=8cm]{./fig05.pdf}
	\end{minipage}
	\hspace{6pt}
	\begin{minipage}[c]{9.8cm}
		\includegraphics[trim= 0 0 2.1cm 0, clip, width=\textwidth]{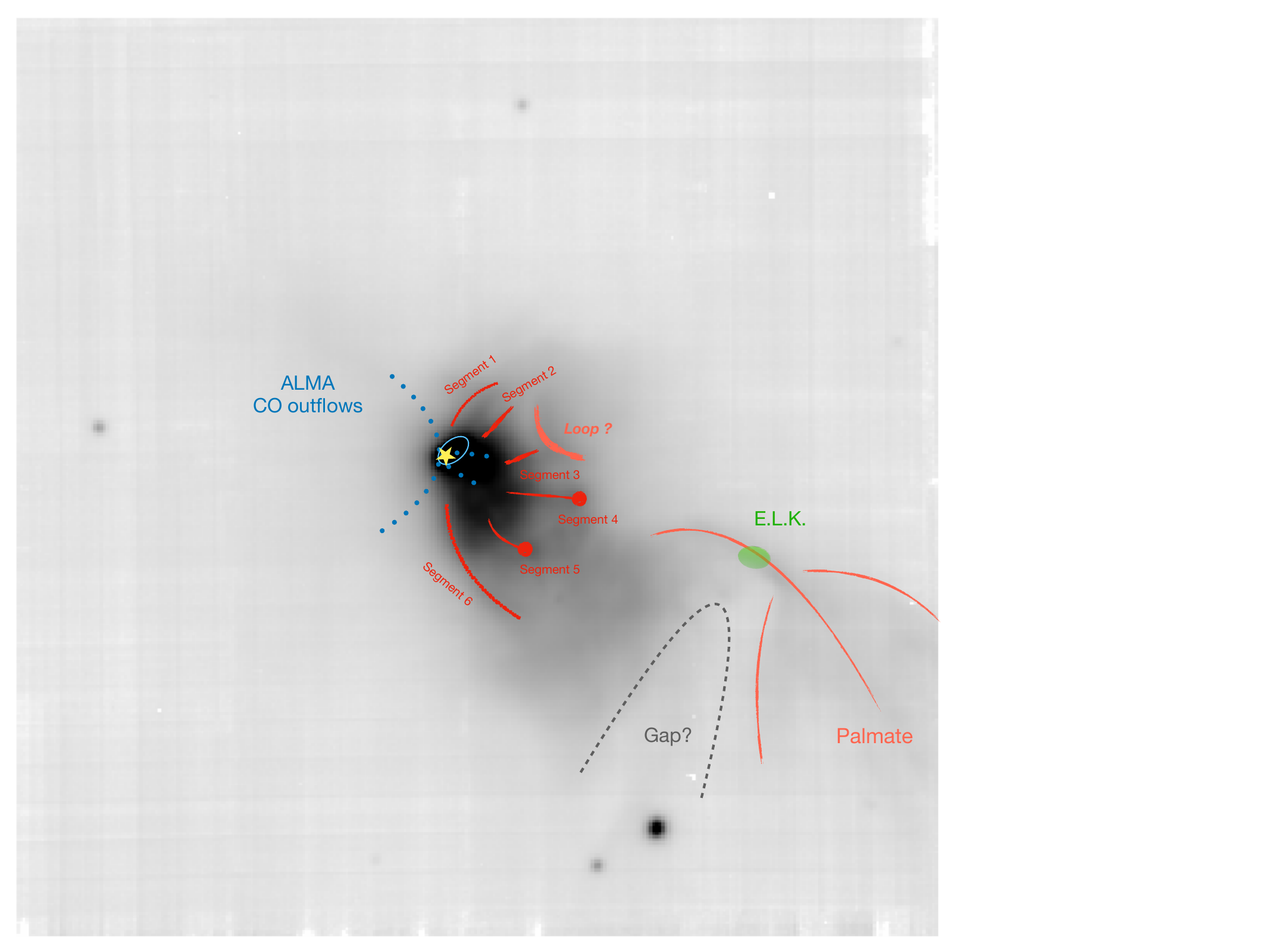}
	\end{minipage}
	\caption{A closer look at the V900 Mon reflection nebula with MUSE/VLT. The {\it left panel} is a reproduction of Fig. \ref{fig:muse_cont_im} to allow direct comparison with the drawing of the individual nebular features in the {\it right panel}. For a description of the features refer to Sect.~\ref{nebmorph}.}
	\label{fig:nebdraw}
\end{figure*}

In Figure~\ref{fig:nebdraw} we attempt to show individual features within the nebula. The bulk of the reflection nebula extends to the west of V900 Mon and to the south-west. Some individual features can be discerned within the nebula depending on the intensity stretch applied to the image. Here, we have opted for an $\sqrt{intensity}$ stretch in both images and have adapted the color palette for clarity: the left panel is a reproduction of Fig. \ref{fig:muse_cont_im}, while the right panel is the same image in black and white. 

We indicatively show the extent of the CO bi-conical outflow (blue dotted lines) detected by ALMA \citep{takami2019}. The location of the star is clearly marked, as is the ellipsoidal extension north-west of the star (cyan), and the emission-line knot (E.L.K; green). There appears to be a gap in the lower-left corner of the nebula (gray, dotted). This has also been noted in Pan-STARRS1 and VPHAS+ broadband images. Within the main bulk of the nebula we can identify six segments (red). Segments 1 and 6 delineate the southern and part of the northern extension of the nebula, while segments 2 to 5 mark features extending outward from the circumstellar region. Segments 4 and 5 appear to terminated at two globules to the west of the star. A loop-like feature is seen at the upper right section of the nebula. The nebula extends further westward like a palmate (i.e., duck's foot) within the MUSE FOV.

It is unclear from the current MUSE data, if individual features expand outward from the circumstellar region. The features could simply be a projection effect within or near the bi-conical outflow. There are no discerning features in individual spectral channels, with the exception of the emission line knot, while a comparison to broadband imaging from the last decade by Pan-STARRS1 and VPHAS+ surveys does not indicate any changes within the nebula. Therefore, we assume that said features have been carved by some earlier mechanism acting in the nebula.

\end{appendix}
\end{document}